%% file: mxmc_20.tex
\documentclass{article}

\usepackage{graphicx}
\usepackage{amsmath}
\usepackage{bm}
\usepackage{amsfonts}
\usepackage{amssymb}
\usepackage{algorithm}
\usepackage{algpseudocode}
\usepackage{book tabs}
\usepackage{color}
\usepackage{subcaption}
\usepackage[table]{xcolor}
\usepackage{rotating}
\usepackage[margin=1.2in]{geometry}
\usepackage{cite}
\usepackage{authblk}
\usepackage{textcomp}
\usepackage{flafter}

\newcommand{\fref}[1]{Figure~\ref{#1}}
\newcommand{\frefs}[2]{Figures~\ref{#1} and \ref{#2}}
\newcommand{\eqs}[2]{Equations~\eqref{#1} and \eqref{#2}}
\newcommand{\sref}[1]{Section~\ref{#1}}
\newcommand{\eref}[1]{Equation~\eqref{#1}}
\newcommand{\tref}[1]{Table~\ref{#1}}

\newcommand{\eg}{\textit{e.g.\ }}

\newcommand{\ie}{{i.e.\ }}

\DeclareMathOperator{\E}{\mathbb{E}}
\DeclareMathOperator{\Var}{\mathbb{V}ar}
\DeclareMathOperator{\Cov}{\mathbb{C}ov}
\renewcommand{\vec}[1]{\underline{#1}}

\newcommand{\PreserveBackslash}[1]{\let\temp=\\#1\let\\=\temp}
\newcolumntype{C}[1]{>{\PreserveBackslash\centering}p{#1}}
\newcolumntype{R}[1]{>{\PreserveBackslash\raggedleft}p{#1}}

\graphicspath{{./figures}}

\title{On the Optimization of Approximate Control Variates with Parametrically Defined Estimators}

\author{G.F. Bomarito}
\author{P.E. Leser}
\author{J.E. Warner}
\author{W.P. Leser}
\affil{National Aeronautics and Space Administration, Langley Research Center, Hampton, VA}

\begin{document}

\maketitle

\begin{abstract}

Multi-model Monte Carlo methods, such as multi-level Monte Carlo (MLMC) and multifidelity Monte Carlo (MFMC),  allow for efficient estimation of the expectation of a quantity of interest given a set of models of varying fidelities.  Recently, it was shown that the MLMC and MFMC estimators are both instances of the approximate control variates (ACV) framework [Gorodetsky et al. 2020]. In that same work, it was also shown that hand-tailored ACV estimators could outperform MLMC and MFMC for a variety of model scenarios. Because there is no reason to believe that these hand-tailored estimators are the best among a myriad of possible ACV estimators, a more general approach to estimator construction is pursued in this work. First, a general form of the ACV estimator variance is formulated. Then, the formulation is utilized to generate parametrically-defined estimators. These parametrically-defined estimators allow for an optimization to be pursued over a larger domain of possible ACV estimators. The parametrically-defined estimators are tested on a large set of model scenarios, and it is found that the broader search domain enabled by parametrically-defined estimators leads to greater variance reduction.

\vspace{10pt}\noindent{\bf Keywords:} Variance reduction; Monte Carlo; Control variates; Multifidelity modeling
\end{abstract}

\input{introduction}

\input{background}
\input{parametric_estimators}

\input{algorithm_summary}

\input{practical_considerations}

\input{literature_examples}

\input{parametric_study}

\input{conclusion}

\section*{Acknowledgements} The authors would like to gratefully acknowledge
funding for this work from two NASA prorams: the Engineering Research \& Analysis Program -- Uncertainty Quantification for Structures \& Dynamics  and the Internal Research and Develoment Program -- Towards Real-Time and High-Precision Trajectory Simulation for Entry, Descent, and Landing Systems

\bibliography{refs}
\bibliographystyle{unsrt}

\input{appendix}

\end{document}

%% file: introduction.tex
\section{Introduction}\label{intro}

Propagating uncertainty through high-fidelity computational models is a cornerstone of computational science and engineering. The most well-known uncertainty propagation method is Monte Carlo (MC) simulation. It can also be considered the most general purpose method because its applicability is independent of the dimensionality of the problem and it produces a fully probabilistic prediction for quantities of interest (QoIs). In practice, however, MC simulation can require tens of thousands of model evaluations and is thus infeasible for expensive models (e.g., those involving discretization of partial differential equations (PDEs)). To reduce computational expense, lower fidelity surrogate models can replace the high-fidelity model within MC simulation, but this yields biased and potentially inaccurate predictions.

As an alternative, multi-model MC methods balance accuracy and computational efficiency by fusing high- and low-fidelity model predictions. A strength of many multi-model approaches is that, similar to MC simulation, their accuracy is independent of the input dimension of the problem. While recent works have sought to match the applicability of MC simulation in terms of the output produced (e.g,. rare event probabilities \cite{peherstorfer_2016, osti_1547423, elfverson_2016}, sensitivity indices \cite{qian_2018}, vector-valued QoIs \cite{QUAGLINO2019300}, probability distributions of QoIs \cite{BIERIG2016661, lu_2016, kennedy_2000, perdikaris_2017, koutsourelakis_2009}), considerable effort has focused on the fundamental problem of constructing accurate multi-model estimators for QoI expectations \cite{Giles08multi-levelmonte, 10.1007/978-3-642-41095-6_4, mfmc2016, gorodetsky_2018, mfmlmc2017, HajiAli2016MultiindexMC}  based on classical control variate (CV) \cite{hesterberg_1996, 10.2307/170564} variance reduction techniques.  Briefly, these techniques leverage the correlation among models to reduce variance in expectation estimates.  Continuing this thread, the current work formulates new CV-based multi-model estimators with improved accuracy.


Generally speaking, multi-model estimators are a linear combination of high-fidelity model evaluations (to ensure unbiasedness) and evaluations of one or more low fidelity models (for computational speedup). Low-fidelity models act as CVs with respect to the high-fidelity model where variance is reduced as correlation among the models increases. Given a fixed computational budget, the crux of multi-model MC is the determination of a sample allocation strategy over available models that minimizes estimator variance. The optimal sample allocation depends on relative model costs and correlations. Optimization requires formulation of estimator variance as an explicit function of sample allocation and the ability to efficiently search the potentially large space of feasible allocations.

Existing efforts in this area are distinguished primarily by the optimization strategy and secondarily by the types of low fidelity models. Two of the most well-known methods, multi-level Monte Carlo (MLMC) \cite{Giles08multi-levelmonte, 10.1007/978-3-642-41095-6_4} and multi-fidelity Monte Carlo (MFMC) \cite{mfmc2016}, both introduce assumptions on model dependency structure that yield closed-form expressions for optimal sample allocation. While MLMC leverages low-fidelity models from a hierarchy of discretization levels and combines model predictions using a telescoping sum (or recursive difference) estimator, MFMC considers arbitrary (e.g., data-driven, reduced-order, analytical) low-fidelity models and combines their predictions using a recursive nested estimator. Simplifying estimator structure in this manner restricts the search for sample allocation to small sub-domains of the entire space, limiting their potential for variance reduction.

Recently, the generalized approximate control variate (ACV) framework \cite{gorodetsky_2018} introduced a unifying formulation for multi-model estimators of which MLMC and MFMC are special cases. By considering a more general model dependency structure and hence a broader search space for sample allocation optimization, a novel set of estimators were designed and shown to substantially improve variance reduction relative to more restricted recursive estimators. Gordetsky et. al also proposed a simple instantiation of \textit{parametrically-defined estimators} called ACVKL. The ACVKL estimator had two parameters that could be used to define multiple sub-domains for optimization, allowing for a more thorough search when iterated exhaustively. However, it was emphasized that superior ACV estimators likely exist and that future work should include more robust numerical optimization schemes.

The work herein may be viewed as a direct extension of \cite{gorodetsky_2018} through theoretical and practical enhancements of the ACV method. In particular, two generalizations of the ACV framework are made to broaden the effective optimization domain and further reduce estimator variance. First, a general expression for estimator variance in terms of sample allocation with \textit{no assumption} on sampling strategy is derived. Second, the full potential of parametrically-defined estimators is explored using tunable recursion, a method for more principled construction and exploration of sub-optimization domains. From these enhancements, three new generalized ACV estimators are defined with demonstrated improvement over their predecessors.

Additionally, this work discusses several practical considerations for increasing robustness and broadening applicability. An automatic model selection procedure is introduced as an additional method for defining optimization subdomains. Implementation details for improving the numerical optimization are also provided. Finally, an open-source Python library \cite{bomarito_2020} is introduced that implements many existing ACV methods, making it easy to reproduce (and extend) the findings in this paper.

It is important to emphasize that this work focuses on \textit{variance reduction} of a MC estimator using the \textit{ACV approach}. As mentioned in \cite{gorodetsky_2018}, this contrasts with mean-squared-error reduction methods with respect to an unknown true model, as provided by MLMC and multi-index MC (MIMC) \cite{HajiAli2016MultiindexMC}. Additionally, the ACV methodology studied here is distinguished from recent work \cite{schaden_2020} that took a fundamentally different approach by formulating a multi-model estimator as a linear regression problem. 
However, some of the practical considerations offered here may be applicable to this alternative approach.

The paper is organized as follows. First, a brief background of the ACV framework is provided in Section 2. Section 3 introduces three new estimators that are generalizations of existing strategies. In Section 4, practical considerations regarding the ACV sample allocation optimization problem are discussed. The performance of the proposed estimators are compared with existing ACV methods on numerical examples in Section 5 before summarizing the work in Section 6.


%% file: background.tex
\section{Background}\label{sec:background}
The parametrically defined estimators developed in the current work are an extension of the ACV framework.  This section recalls the necessary fundamentals of the ACV framework from the work of Gorodetsky et al.\cite{gorodetsky_2018}; extensions of the framework are explored in the subsequent sections.

\subsection{Approximate Control Variate (ACV) Estimation}
Let $\{Q_0, Q_1, ..., Q_M\}$ be a set of $M + 1$ mappings $Q_i : \mathbb{R}^d \rightarrow \mathcal{Q}_i \subset  \mathbb{R}$ from vector-valued inputs to scalar outputs.  These mappings will henceforth be referred to as models. The primary goal of ACVs is to estimate the expectation of $Q_0$ (\ie $\mu_0 = \E[Q_0]$) using a set of $N$ samples $z = \{z^{(0)}, z^{(1)}, ... z^{(N)} \}$ of the input random variables and $M$ correlated models $\{Q_1, Q_2..., Q_M\}$.  Since the $Q_0$ model holds particular significance it is referred to as the high-fidelity model, the remaining models are referred to as low-fidelity models.  The form of the ACV estimator is\footnote{As in the referenced work, attention is restricted to the linear control variate case.}:
\begin{equation}
\tilde {Q}(\vec{\alpha}, z, \mathcal{A}) = \hat Q_0 (z_0) + \sum_{i=1}^M\alpha_i \left( \hat Q_i (z^*_i) -  \hat Q_i (z_i) \right) = \hat Q_0 + \vec{\alpha}^T \vec{\Delta}
\label{eq:acv}
\end{equation}
where $\vec{\alpha} = \{ \alpha_1, \alpha_2, .., \alpha_M \} $ are the control variate weights and $\hat Q_k(z_\ell)$ is the MC estimator for $\E[Q_k]$ using samples $z_\ell$.  In the vectorized form of the equation on the right side $\vec{\Delta} = \{ \Delta_1(z^*_1, z_1), ..., \Delta_M(z^*_M, z_M) \}$ and $\Delta_i(z^*_i, z_i) =\hat Q_i (z^*_i) -  \hat Q_i (z_i)$.  Note that the explicit dependence $\vec{\Delta}$ on $z^*_i, z_i$ has been dropped for simplification of notation.
In the above equation, $z_0$, $z_i$, and $z^*_i$ are subsets of $z$ such that their union is $z$: \ie $z_0 \cup z^*_1 \cup z_1 \cup ... \cup z^*_M \cup z_M = z$.  The subsets can, and usually do, overlap such that $z_i, z^*_i \cap z_j, z^*_j \neq \varnothing$.  $\mathcal{A}$ represents the information associated with the allocation of $z$ into its subsets $z_0$, $z_i$, and $z^*_i$.  Specifically, $\mathcal{A}$ contains the number of samples in $z$, the number of samples in each subset, the number of samples in the intersections of pairs of subsets, and the number of samples in the union of pairs of subsets: $N$, $N_i$, $N_{i\cap j}$, and $N_{i\cup j}$, respectively.  

Because \eref{eq:acv} is an unbiased estimator, the minimum variance estimator is the one with minimal error.  
The variance of the ACV estimator is:
\begin{equation}
\Var[\tilde Q(\vec{\alpha}, z, \mathcal{A})] = \Var[\hat Q_0] +\vec{\alpha}^T(\Cov[\vec{\Delta}, \vec{\Delta}]) \vec{\alpha} + 2\vec{\alpha}^T \Cov[\vec{\Delta}, \hat Q_0]
 \label{eq:acv_est_var}
\end{equation}
Note that again the explicit dependence on $z^*_i$ and $z_i$ of the terms on the right hand side have been dropped for simplification of notation.
The optimal control variate weights, corresponding to a minimum variance for a given $z$ and $\mathcal{A}$ are then
\begin{align}
\begin{split}
 \vec{\alpha}_{opt}(z, \mathcal{A}) &= \arg \min_{\vec{\alpha}} \Var[\tilde {Q}(\vec{\alpha}, z, \mathcal{A})] = -\Cov[\vec{\Delta}, \vec{\Delta}]^{-1}\Cov[\vec{\Delta}, \hat Q_0]
 \end{split}
 \label{eq:opt_alpha}
\end{align}
Inserting \eref{eq:opt_alpha} into \eref{eq:acv_est_var} gives the variance of the $\alpha$-optimal ACV estimator:
\begin{equation}
\Var[\tilde Q^{\vec{\alpha}_{opt}}(z, \mathcal{A})] = \Var[\hat Q_0] - \Cov[\vec{\Delta}, \hat Q_0]^T \Cov[\vec{\Delta}, \vec{\Delta}]^{-1} \Cov[\vec{\Delta}, \hat Q_0]
 \label{eq:acv_est_var_alpha_optimal}
\end{equation}

\subsection{ACV Variance as a Function of Allocation}

\begin{figure}
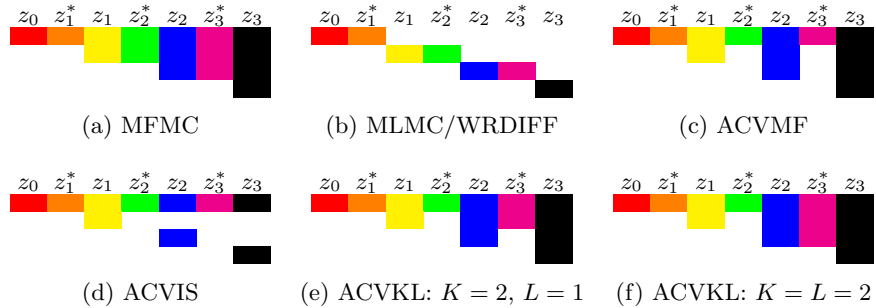

\centering
\small
\setlength\tabcolsep{1.5pt}
\renewcommand{\arraystretch}{0.6}
\begin{subfigure}{12em}
  \centering
\begin{tabular}{C{1.2em}C{1.2em}C{1.2em}C{1.2em}C{1.2em}C{1.2em}C{1.2em}}
$z_0$ & $z^*_1$ & $z_1$& $z^*_2$ & $z_2$& $z^*_3$ & $z_3$ \\
\cellcolor{red}&\cellcolor{orange}&\cellcolor{yellow}&\cellcolor{green}&\cellcolor{blue}&\cellcolor{magenta}&\cellcolor{black}\\
&&\cellcolor{yellow}&\cellcolor{green}&\cellcolor{blue}&\cellcolor{magenta}&\cellcolor{black}\\
&&&&\cellcolor{blue}&\cellcolor{magenta}&\cellcolor{black}\\
&&&&&&\cellcolor{black}
\end{tabular}
\caption{MFMC}
\label{fig:mfmc_allocation}
\end{subfigure}
\begin{subfigure}{12em}
  \centering
\begin{tabular}{C{1.2em}C{1.2em}C{1.2em}C{1.2em}C{1.2em}C{1.2em}C{1.2em}}
$z_0$ & $z^*_1$ & $z_1$& $z^*_2$ & $z_2$& $z^*_3$ & $z_3$ \\
\cellcolor{red}&\cellcolor{orange}&&&&&\\
&&\cellcolor{yellow}&\cellcolor{green}&&&\\
&&&&\cellcolor{blue}&\cellcolor{magenta}&\\
&&&&&&\cellcolor{black}
\end{tabular}
\caption{MLMC/WRDIFF}
\label{fig:mlmc_and_wrdiff_allocation}
\end{subfigure}
\begin{subfigure}{12em}
  \centering
\begin{tabular}{C{1.2em}C{1.2em}C{1.2em}C{1.2em}C{1.2em}C{1.2em}C{1.2em}}
$z_0$ & $z^*_1$ & $z_1$& $z^*_2$ & $z_2$& $z^*_3$ & $z_3$ \\
\cellcolor{red}&\cellcolor{orange}&\cellcolor{yellow}&\cellcolor{green}&\cellcolor{blue}&\cellcolor{magenta}&\cellcolor{black}\\
&&\cellcolor{yellow}&&\cellcolor{blue}&&\cellcolor{black}\\
&&&&\cellcolor{blue}&&\cellcolor{black}\\
&&&&&&\cellcolor{black}
\end{tabular}
\caption{ACVMF}
\label{fig:acvmf_allocation}
\end{subfigure}
\begin{subfigure}{12em}
\vspace{10pt}
  \centering
\begin{tabular}{C{1.2em}C{1.2em}C{1.2em}C{1.2em}C{1.2em}C{1.2em}C{1.2em}}
$z_0$ & $z^*_1$ & $z_1$& $z^*_2$ & $z_2$& $z^*_3$ & $z_3$ \\
\cellcolor{red}&\cellcolor{orange}&\cellcolor{yellow}&\cellcolor{green}&\cellcolor{blue}&\cellcolor{magenta}&\cellcolor{black}\\
&&\cellcolor{yellow}&&&&\\
&&&&\cellcolor{blue}&&\\
&&&&&&\cellcolor{black}
\end{tabular}
\caption{ACVIS}
\label{fig:acvis_allocation}
\end{subfigure}
\begin{subfigure}{12em}
\vspace{10pt}
  \centering
\begin{tabular}{C{1.2em}C{1.2em}C{1.2em}C{1.2em}C{1.2em}C{1.2em}C{1.2em}}
$z_0$ & $z^*_1$ & $z_1$& $z^*_2$ & $z_2$& $z^*_3$ & $z_3$ \\
\cellcolor{red}&\cellcolor{orange}&\cellcolor{yellow}&\cellcolor{green}&\cellcolor{blue}&\cellcolor{magenta}&\cellcolor{black}\\
&&\cellcolor{yellow}&&\cellcolor{blue}&\cellcolor{magenta}&\cellcolor{black}\\
&&&&\cellcolor{blue}&&\cellcolor{black}\\
&&&&&&\cellcolor{black}
\end{tabular}
\caption{ACVKL: $K=2$, $L=1$}
\label{fig:acvkl_2_1_allocation}
\end{subfigure}
\begin{subfigure}{12em}
\vspace{10pt}
  \centering
\begin{tabular}{C{1.2em}C{1.2em}C{1.2em}C{1.2em}C{1.2em}C{1.2em}C{1.2em}}
$z_0$ & $z^*_1$ & $z_1$& $z^*_2$ & $z_2$& $z^*_3$ & $z_3$ \\
\cellcolor{red}&\cellcolor{orange}&\cellcolor{yellow}&\cellcolor{green}&\cellcolor{blue}&\cellcolor{magenta}&\cellcolor{black}\\
&&\cellcolor{yellow}&&\cellcolor{blue}&\cellcolor{magenta}&\cellcolor{black}\\
&&&&\cellcolor{blue}&\cellcolor{magenta}&\cellcolor{black}\\
&&&&&&\cellcolor{black}
\end{tabular}
\caption{ACVKL: $K=L=2$}
\label{fig:acvkl_2_2_allocation}
\end{subfigure}
\caption{Schematics of sample allocations for four models ($M=3$) with different sample allocation strategies.  In each schematic, the columns/colors represent the seven subsets of $z$ required to define a sample allocation for four models.  Colors that share the same row indicate that a group of samples are shared between those subsets.}
\label{fig:lit_allocations}
\end{figure}

The Multi-model MC estimators associated with MFMC and MLMC have been shown to be specific instances of ACV estimators, each implementing a different sample allocation strategy \cite{gorodetsky_2018}.  The MFMC estimator implements a recursive nested sample allocation strategy (\fref{fig:mfmc_allocation}) in which $z_i \subset z_j$ for $i <j$ and $z^*_i = z_{i-1}$.  The MLMC estimator implements a recursive difference sample allocation (\fref{fig:mlmc_and_wrdiff_allocation}) in which $z^*_i = z_{i-1}$ and all $z_i$ are disjoint (\ie independent).  In contrast to the other ACV estmators, the MLMC estimator assumes fixed values of $\alpha_i = -1$.  This assumption yields simple closed-form expressions for sample allocation and facilitates analytical error analysis \cite{Giles08multi-levelmonte, 10.1007/978-3-642-41095-6_4} but at the cost of a generally non-optimal estimator \cite{gorodetsky_2018}.    The generalization of the MLMC sample allocation strategy to include optimal $\vec{\alpha}$ was dubbed the weighted recursive difference (WRDIFF) method by Gorodetsky et al. \cite{gorodetsky_2018}.

Gorodetsky et al. \cite{gorodetsky_2018} introduced two more sample allocation strategies. The strategy named ACVMF is based on the multifidelity sampling strategy with $z_i \subset z_j$ for $i <j$ and $z^*_i = z_0$,  \fref{fig:acvmf_allocation}.  The strategy named ACVIS uses an independent sampling method with $z_i \cap z_j = z_0$ and $z^*_i = z_0$,  \fref{fig:acvis_allocation}.  Both of these strategies have improved variance-reduction potential because all low-fidelity models act as control variates of the high fidelity model, \ie each $\Delta_i$ is highly correlated with $\hat{Q_0}$.

For the above listed methods, it has been shown that the variance of the ACV estimator and $\vec{\alpha}_{opt}$ do not depend on $z$ rather only on the sample allocation $\mathcal{A}$ and the model statistics \cite{gorodetsky_2018}.  Thus, assuming fixed model statistics, the variance of these ACV estimators becomes solely a function of sample allocation: 
\begin{equation}
\Var[\tilde{Q}^{\vec{\alpha}_{opt}}](\mathcal{A}) = \Var[\tilde Q^{\vec{\alpha}_{opt}}(z, \mathcal{A})]
\end{equation}
This is true in general (as is shown in \sref{sec:general_acv_variance}), but has of yet only been illustrated for specific sample allocation strategies.  For instance, the variances of the ACVMF and ACVIS estimators have been related to $\mathcal{A}$ through the individual components of \eref{eq:acv_est_var_alpha_optimal}: the scalar $\Var[\hat Q_0]$, the matrix $\Cov[\vec{\Delta}, \vec{\Delta}]$, and the vector $\Cov[\vec{\Delta}, \hat Q_0]$.  The first of these terms is related to $\mathcal{A}$ as $\Var[\hat Q_0] = N_0^{-1}\Var[Q_0]$.  The remaining two terms are linked to $\mathcal{A}$ in the following form:
\begin{align}
\Cov[\vec{\Delta}, \vec{\Delta}] &= \mathbf{F}(\mathcal{A}) \circ \mathbf{C} \label{eq:cov_d_d_F}\\
\Cov[\vec{\Delta}, \hat Q_0] &= \text{diag}(\mathbf{F}(\mathcal{A})) \circ \mathbf{c} \label{eq:cov_d_0_F}
\end{align}
where $\mathbf{C} \in \mathbb{R}^{M\text{x}M}$ is the covariance matrix among low-fidelity models $\{Q_1, ..., Q_M\}$ and $\mathbf{c} \in \mathbb{R}^M$ is the covariance vector of the low fidelity models $\{Q_1, ..., Q_M\}$ with the high fidelity model $Q_0$.  The $\circ$ operator signifies the Hadamard (elementwise) product.  The matrix $\mathbf{F} \in \mathbb{R}^{M\text{x}M}$ is a function of $\mathcal{A}$ specific to the sample allocation strategy.

\subsection{ACV Optimization} 
Let $\vec{w} = \{w_0, ..., w_M\} $ represent the costs associated with the evaluation of each of the models.  The total cost of the ACV estimator is:
\begin{equation}
\tilde{w}(\vec{w}, \mathcal{A}) = w_0N_0 + \sum_{i=1}^M w_iN_{i^* \cup i}
\end{equation}

The formal definition of the ACV optimization problem is here defined as the common case where an estimator for $\E[Q_0]$ is sought with minimum variance given a set of models $\{Q_0, ..., Q_M\}$, costs, $\vec{w}$, and a constraint on the total cost $\tilde{w}^*$. This simplifies to the identification of an optimal sample allocation $\mathcal{A}$:
\begin{equation}
\begin{split}
\min_{\mathcal{A}\in\mathbb{A}} & \Var[\tilde{Q}^{\vec{\alpha}_{opt}}](\mathcal{A})\\
\text{s.t.} &~\tilde{w}(\vec{w}, \mathcal{A}) \leq \tilde{w}^*
\end{split}
\label{eq:main_opt}
\end{equation}
where $\mathbb{A}$ is the set of all possible sample allocations.  Since $\mathcal{A}$ includes not only the number of samples in $z$ but also its relative division into $2M + 1$ subsets, the domain of the optimization $\mathbb{A}$ is astronomically large.  In addition to the difficulty caused by the size of $\mathbb{A}$, an efficient means of parameterizing the domain is lacking.  This is compounded by the fact that calculation of $\vec{\alpha}_{opt}$ precludes a general analytical formula for the estimator variance.  

All of the previously developed ACV methods approach the optimization in \eref{eq:main_opt} by restricting the optimization to subsets of $\mathbb{A}$ using an assumed sample allocation strategy\footnote{Standard MLMC also assumes values of $\alpha_i=-1$, which is generally different than the $\vec{\alpha}_{opt}$.}.  The sub-domains are smaller and easier to parameterize than $\mathbb{A}$; in the case of MLMC and MFMC the sub-domains are restricted sufficiently to permit analytical solutions of the optimization.  A key consequence of restricting optimization to sub-domains of $\mathbb{A}$ is that the optimal values found on the restricted domain are local minima and may not reflect the global minimum over $\mathbb{A}$.

A very practical approach to the optimization is to break the problem into several sub-optimization problems.  The simplest case of performing the sub-optimization approach, which is common in practice, is to calculate the variance of multiple estimators like MLMC and MFMC then use the method with lowest variance.  Formally, the sub-optimization approach is
\begin{equation}
\min_{A_p \in \{1, \dots, P\}} \left(\begin{array}{rl}
\min\limits_{\mathcal{A}\in A_p} & \Var[\tilde{Q}^{\vec{\alpha}_{opt}}](\mathcal{A})\\
\text{s.t.} &~\tilde{w}(\vec{w}, \mathcal{A}) \leq \tilde{w}^*\\
\end{array}\right)
\label{eq:sub_opts}
\end{equation}
where $A_p \subset \mathbb{A}$ is one of a set of $P$ sub-domains.  As the effective optimization domain approaches $\mathbb{A}$ (\ie $A_1 \cup \dots \cup A_P \rightarrow \mathbb{A}$) the result of \eref{eq:sub_opts} approaches the result of \eref{eq:main_opt}; this of course assumes that the sub-optimizations find global minima in their restricted domains. 

With the intent of broadening the effective optimization domain, a parametrically-defined estimator, coined ACVKL \cite{gorodetsky_2018}, was developed that procedurally defines sub-domains of $\mathbb{A}$.  This corresponds to a parametric definition of the sample allocation strategy, in this case by two parameters $K \in \{1,\dots,M\}$ and $L \in \{0,\dots,K\}$.  The ACVKL sample allocation strategy, shown in \frefs{fig:acvkl_2_1_allocation}{fig:acvkl_2_2_allocation}, is defined by: $z_i \subset z_j$ for $i < j$, $z^*_i = z_0$ for $i \le K$, and $z^*_i = z_L$ for $i > K$.  

%% file: parametric_estimators.tex
\section{Parametrically-defined Estimators} \label{sec:parametric_estimators}
There is no reason to believe that sub-domains covered by the methods described in \sref{sec:background} capture the global minimum. Therefore, the intent of the current work is to enable the broadening of the effective optimization domain as much as possible through the introduction of parametrically-defined estimators. In this section, three such estimators are developed based on the derivation of a general expression for the variance of the ACV estimator as a function of sample allocation.  The three estimators are generalizations of sampling strategies in the literature produced using tunable recursion. Further generalizations are introduced through relaxation of ordering constraints and an automatic model selection scheme. All of these concepts are discussed in detail in the following subsections.

\subsection{The General Expression for the Dependence of ACV Variance on Sample Allocation} \label{sec:general_acv_variance}
Again assuming fixed model statistics, the general expression for ACV variance as solely a function of sample allocation, \ie $\Var[\tilde Q](\mathcal{A})$, is sought.  The motivation here is that the construction of the expression will enable more general parametric definitions of sample allocation strategies.  Recall first that the general equation for ACV variance in \eref{eq:acv_est_var} is a function of $\Var[\hat Q_0]$, $\vec{\alpha}$, $\Cov[\vec{\Delta}, \vec{\Delta}]$, and $\Cov[\vec{\Delta}, \hat Q_0]$.  Thus, it suffices to show a general expression of each of these quantities as a function of sample allocation.  $\Var[\hat Q_0] = N_0^{-1}\Var[Q_0]$ is a function of sample allocation through its reliance on $N_0$.  Because optimal $\vec{\alpha}$ are a function of $\Cov[\vec{\Delta}, \vec{\Delta}]$ and $\Cov[\vec{\Delta}, \hat Q_0]$, its dependence on $\mathcal{A}$ is illustrated through their dependence on $\mathcal{A}$\footnote{An assumption of a constant $\vec{\alpha}$, such as $\alpha_i = -1$ in the MLMC estimator, could be used here and does not change the fact that ACV estimator variance can be expressed solely as a function of sample allocation.}

The general relationship between $\Cov[\vec{\Delta}, \vec{\Delta}]$, $\Cov[\vec{\Delta}, \hat Q_0]$, and $\mathcal{A}$, as derived in the appendix, is:
\begin{align}
\Cov[\vec{\Delta}, \vec{\Delta}] &= \mathbf{G}(\mathcal{A}) \circ \mathbf{C} \label{eq:cov_delta_delta}\\
\Cov[\vec{\Delta}, \hat Q_0] &= \mathbf{g}(\mathcal{A}) \circ \mathbf{c} \label{eq:cov_delta_0}
\end{align}
with
\begin{align}
G_{ij} & = \frac{N_{i* \cap j*}}{N_{i*}N_{j*}} - \frac{N_{i* \cap j}}{N_{i*}N_j}  - \frac{N_{i \cap j*}}{N_{i}N_{j*}}+ \frac{N_{i \cap j}}{N_{i}N_j} \\
g_{i} & = \frac{N_{i* \cap 0}}{N_{i*}N_0} - \frac{N_{i \cap 0}}{N_{i}N_0}
\end{align}
Note that the forms of \eqs{eq:cov_d_d_F}{eq:cov_d_0_F} are similar to the above, and represent special cases when $\mathbf{g} = \text{diag}(\mathbf{G})$, such as cases like ACVMF and ACVIS, where $z^*_i = z_0$ and $z^*_i \subset z_i$.

\subsection{Tunable Recursion}
The ACVKL estimator is based on tunable recursion: \ie that the recursion structure of the control variate scheme can be parametrically adjusted.  In the ACVKL estimator, the first $K$ low-fidelity models act as control variates for the high-fidelity model, and the remaining models act as control variates for estimating the control mean of model $L$.  In this context, acting as a control variate implies that the first MC estimator of a model is maximally correlated with its target, \ie it shares the same set of samples.  More formally, if model $i$ is to act as a control variate for model $j$  then $z^*_i = z_j$.  

\begin{figure}
\centering
\includegraphics[scale=0.5]{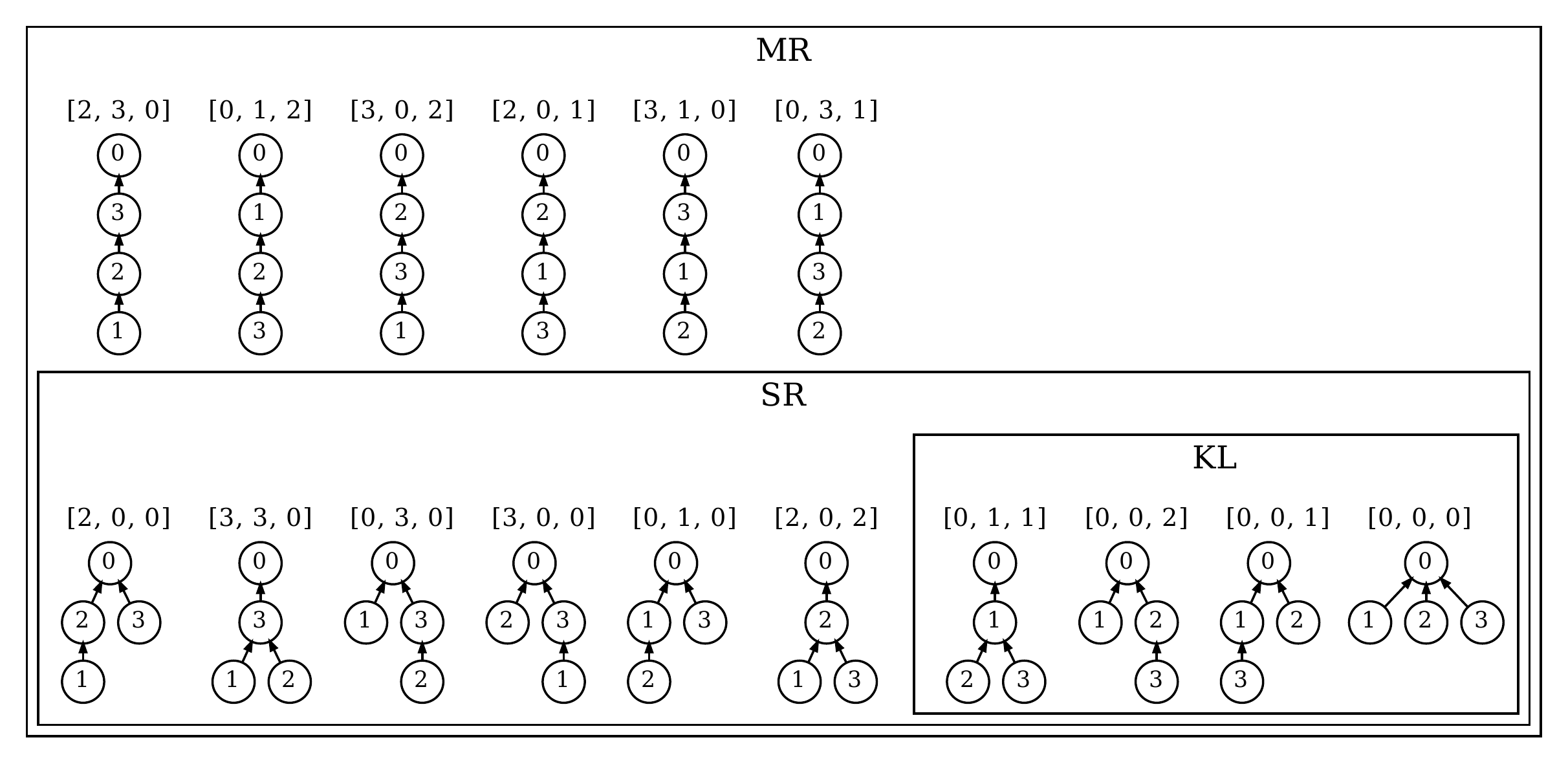}
\caption{All possible recusion trees with four models. $\raisebox{.5pt}{\textcircled{\raisebox{-.4pt} {\scriptsize $i$}}}\leftarrow$ \raisebox{.5pt}{\textcircled{\raisebox{-.4pt} {\scriptsize $j$}}} indicates that model $j$ acts as a control variate for model $i$.  The boxes represent the MR, SR, and KL recursion strategies.  Note that KL $\subset$ SR $\subset$ MR.  The list of integers above each tree represents $\beta$ for that case.}
\label{fig:recursion_trees}
\end{figure}

\begin{figure}
\centering
\includegraphics[scale=1.0]{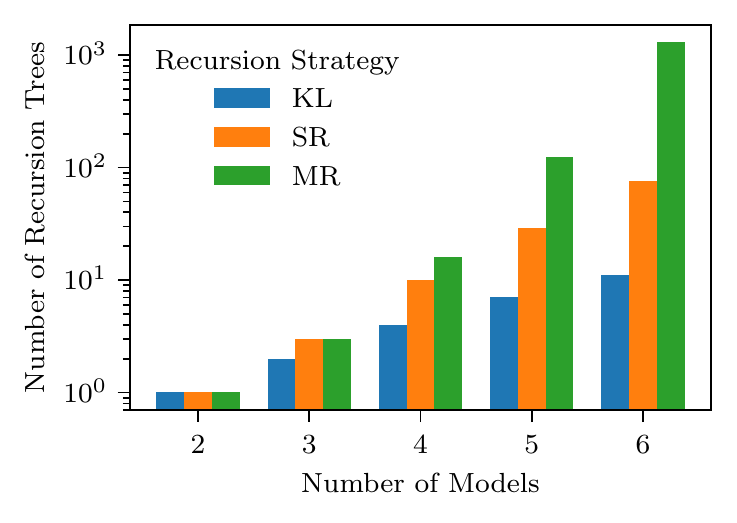}
\caption{Number of recursion trees as a function of the number of models.  The MR recursion strategy corresponds to all possible recursion trees.}
\label{fig:num_recursions}
\end{figure}

The notion of tunable recursion can be generalized further to include all possible ways of assigning models to act as control variates for other models.  Given that each low-fidelity model may act as a control variate for another model and that the goal is to reduce variance of the high-fidelity model, then there exists a zero-rooted acyclic tree which represents its recursion structure.  \fref{fig:recursion_trees} illustrates all the possible recursion structures for the case of four models, wherein each model (represented as a node in the tree) is connected by a directed edge signifying a control variate relationship.  Each recursion tree (and a corresponding sample allocation strategy) defines a sub-domain in $\mathbb{A}$ upon which to perform a sub-optimization.  Thus, when used in a sub-optimization scheme, the number of recursion trees enumerated relates to the size of the effective domain of the optimization.  The number of enumerated recursion trees also relates directly to the computational cost of performing the overall optimization.  This is not a reason for concern in most cases, especially since the sub-optimizations may be performed in parallel; however, due to the combinatorial nature of the recursion trees, the number of possible trees increases rapidly with increased number of models (see \fref{fig:num_recursions}).   For this reason, it may be worth considering more limited recursion strategies.

The KL recursion strategy discussed above limits the recursion trees to those with a depth less than or equal to two, in which model $i$ can be a control variate of model $j$ if $i> j$.  A slightly less restricted version of the KL recursion strategy contains all recursion trees with a depth less than or equal to 2, and assumes nothing about the ordering; this recursion strategy is henceforth called SR for single recursion.  The recursion strategy which contains all possible recursion trees regardless of size is henceforth called MR for multiple recursion.  Each of these recursion strategies is illustrated as a box in \fref{fig:recursion_trees}.  \fref{fig:num_recursions} shows that adding restrictions to the recursion trees drastically slows the explosion of possibilities with increasing numbers of models.

Every recursion tree can be represented by a mapping $\vec{\beta} \in \mathbb{R}^{M}$ from each low-fidelity model to the target of its control variate estimation, \ie $\beta_i = j$ implies $z^*_i = z_j$.  Note that the index of the $\vec{\beta}$ vector starts at 1, $i \in \{1, ..., M\}$, and contains values $j \in \{0, ..., M\}$.  As an example, the KL recursion strategy can be described with:
\begin{equation}
\beta^{KL} _i = \left\{\begin{array}{cl}
0 &: i \leq K\\
L &: i > K \\
\end{array}\right.
\end{equation}
The $\beta$ vector is illustrated for all the four-model recursion trees in \fref{fig:recursion_trees}.

The last two ingredients to a tunable recursion sub-optimization are the development of (1) a variance calculation based on the recursion structure and (2) a means to calculate the number of model evaluations $N_0, N_{i^* \cup i}$ for the total cost calculation.  

\subsubsection{Generalized Multifidelity (GMF) Estimators}
The multifidelity sample allocation strategy used in MFMC and ACVMF are used here to create a generalized multifidelity estimator.  In this sample allocation strategy, all subsets of $z$ are defined based on their size, $N_i$, and include the first $N_i$ samples of $z$: \ie $z_i = \{z^{(1)}, \dots, z^{(N_i)} \}$.  

Based on the multifidelity sample allocation strategy, where $N{i\cap j} = \min(N_i, N_j)$, and the definition of $\beta$, the elements of $\mathbf{G}$ and $\mathbf{g}$ are defined as follows:
\begin{equation}
\begin{split}
G_{ij}^{GMF} 
& = \frac{\min(N_{\beta_i}, N_{\beta_j})}{N_{\beta_i}N_{\beta_j}} - \frac{\min(N_{\beta_i}, N_j)}{N_{\beta_i}N_j}  - \frac{\min(N_i, N_{\beta_j})}{N_{i}N_{\beta_j}} + \frac{\min(N_i,  N_j)}{N_{i}N_j} 
\end{split}
\end{equation}
\begin{equation}
\begin{split}
g_{i}^{GMF} 
& = \frac{\min(N_{\beta_i},N_0)}{N_{\beta_i}N_0} - \frac{\min(N_i, N_0)}{N_{i}N_0}\\
\end{split}
\end{equation}
From here, \eqs{eq:cov_delta_delta}{eq:cov_delta_0} can be used in conjunction with \eref{eq:acv_est_var_alpha_optimal} to produce the generalized multifidelity (GMF) estimator variance, $\Var[\tilde{Q}^{GMF}]$, as a function of $\vec{N} = \{N_0, \dots, N_M\}$ and $\vec{\beta}$.

Analogously, the number of model evaluations are $N_{i^* \cup i} = \max(N_{\beta_i}, N_i)$ for the GMF estimator.  Again noting that they are a function of $\vec{N}$ and $\vec{\beta}$.

The resulting sub-optimization scheme is
\begin{equation}
\min_{\vec{\beta}} \left(\begin{array}{rl}
\min\limits_{\vec{N}} & \Var[\tilde{Q}^{GMF}](\vec{N}, \vec{\beta})\\
\text{s.t.} &~\tilde{w}(\vec{N}, \vec{\beta}) \leq \tilde{w}^*\\
\end{array}\right)
\end{equation}
Where each sub-optimization is performed on variables $\vec{N}$ associated with the sizes of the subsets of $z$.

\begin{figure}
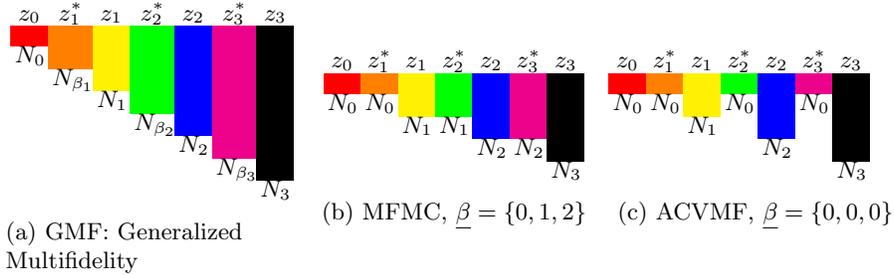

\centering
\small
\setlength\tabcolsep{1.5pt}
\renewcommand{\arraystretch}{0.7}

\begin{subfigure}{12em}
\centering
\begin{tabular}{C{1.2em}C{1.5em}C{1.2em}C{1.5em}C{1.2em}C{1.5em}C{1.2em}}
$z_0$ & $z^*_1$ & $z_1$& $z^*_2$ & $z_2$& $z^*_3$ & $z_3$ \\
\cellcolor{red}&\cellcolor{orange}&\cellcolor{yellow}&\cellcolor{green}&\cellcolor{blue}&\cellcolor{magenta}&\cellcolor{black}\\
$N_0$&\cellcolor{orange}&\cellcolor{yellow}&\cellcolor{green}&\cellcolor{blue}&\cellcolor{magenta}&\cellcolor{black}\\
&$N_{\beta_1}$&\cellcolor{yellow}&\cellcolor{green}&\cellcolor{blue}&\cellcolor{magenta}&\cellcolor{black}\\
&&$N_1$&\cellcolor{green}&\cellcolor{blue}&\cellcolor{magenta}&\cellcolor{black}\\
&&&$N_{\beta_2}$&\cellcolor{blue}&\cellcolor{magenta}&\cellcolor{black}\\
&&&&$N_2$&\cellcolor{magenta}&\cellcolor{black}\\
&&&&&$N_{\beta_3}$&\cellcolor{black}\\
&&&&&&$N_3$
\end{tabular}
\caption{\raggedright GMF: Generalized Multifidelity}
\end{subfigure}
\begin{subfigure}{12em}
  \centering
\begin{tabular}{C{1.2em}C{1.2em}C{1.2em}C{1.2em}C{1.2em}C{1.2em}C{1.2em}}
$z_0$ & $z^*_1$ & $z_1$& $z^*_2$ & $z_2$& $z^*_3$ & $z_3$ \\
\cellcolor{red}&\cellcolor{orange}&\cellcolor{yellow}&\cellcolor{green}&\cellcolor{blue}&\cellcolor{magenta}&\cellcolor{black}\\
$N_0$&$N_0$&\cellcolor{yellow}&\cellcolor{green}&\cellcolor{blue}&\cellcolor{magenta}&\cellcolor{black}\\
&&$N_1$&$N_1$&\cellcolor{blue}&\cellcolor{magenta}&\cellcolor{black}\\
&&&&$N_2$&$N_2$&\cellcolor{black}\\
&&&&&&$N_3$
\end{tabular}

\caption{MFMC, $\vec{\beta} = \{0, 1, 2 \}$}
\end{subfigure}
\begin{subfigure}{12em}
\begin{tabular}{C{1.2em}C{1.2em}C{1.2em}C{1.2em}C{1.2em}C{1.2em}C{1.2em}}
$z_0$ & $z^*_1$ & $z_1$& $z^*_2$ & $z_2$& $z^*_3$ & $z_3$ \\
\cellcolor{red}&\cellcolor{orange}&\cellcolor{yellow}&\cellcolor{green}&\cellcolor{blue}&\cellcolor{magenta}&\cellcolor{black}\\
$N_0$&$N_0$&\cellcolor{yellow}&$N_0$&\cellcolor{blue}&$N_0$&\cellcolor{black}\\
&&$N_1$&&\cellcolor{blue}&&\cellcolor{black}\\
&&&&$N_2$&&\cellcolor{black}\\
&&&&&&$N_3$
\end{tabular}
\caption{ACVMF, $\vec{\beta} = \{0, 0, 0 \}$}
\end{subfigure}
\caption[]{Generalized multifidelity sample allocations for four models\footnotemark. Labels below the column indicate the number of samples.}
\hrule width 0.4\textwidth
\raggedright
\vspace{2pt}\small\textsuperscript{4} See \fref{fig:lit_allocations} for further description of sample allocation schematics.
\label{fig:gmf_diagram}
\end{figure}

A schematic view of the sample allocation associated with the GMF estimator is shown in \fref{fig:gmf_diagram}.  It also illustrates how MFMC and ACVMF are equivalent to GMF with specific values of $\vec{\beta}$.  The $\vec{\beta}$ associated with fully recursive structure of MFMC is $\{0, 1,\dots, M-1\}$.  The $\vec{\beta}$ associated with ACVMF is $\vec{0}$.

\subsubsection{Generalized Recursive Difference (GRD) Estimators}
\begin{figure}
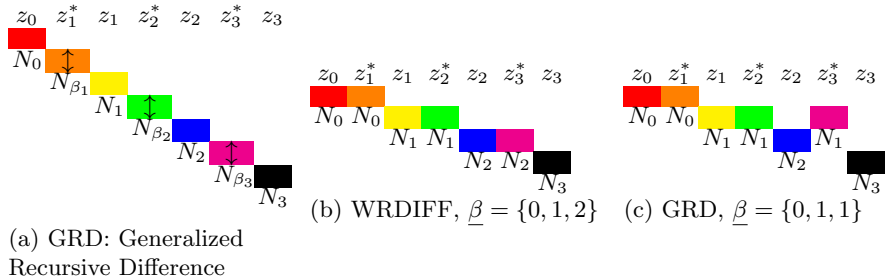

\centering
\small
\setlength\tabcolsep{1.5pt}
\renewcommand{\arraystretch}{0.7}

\begin{subfigure}{12em}
\begin{tabular}{C{1.2em}C{1.5em}C{1.2em}C{1.5em}C{1.2em}C{1.5em}C{1.2em}}
$z_0$ & $z^*_1$ & $z_1$& $z^*_2$ & $z_2$& $z^*_3$ & $z_3$ \\
\cellcolor{red}&&&&&&\\
$N_0$&\cellcolor{orange} $\updownarrow$&&&&&\\
&$N_{\beta_1}$&\cellcolor{yellow}&&&&\\
&&$N_1$&\cellcolor{green}$\updownarrow$&&&\\
&&&$N_{\beta_2}$&\cellcolor{blue}&&\\
&&&&$N_2$&\cellcolor{magenta}$\updownarrow$&\\
&&&&&$N_{\beta_3}$&\cellcolor{black}\\
&&&&&&$N_3$
\end{tabular}
\caption{\raggedright GRD: Generalized \linebreak Recursive Difference}
\end{subfigure}
\begin{subfigure}{12.5em}
\begin{tabular}{C{1.2em}C{1.2em}C{1.2em}C{1.2em}C{1.2em}C{1.2em}C{1.2em}}
$z_0$ & $z^*_1$ & $z_1$& $z^*_2$ & $z_2$& $z^*_3$ & $z_3$ \\
\cellcolor{red}&\cellcolor{orange}&&&&&\\
$N_0$&$N_0$&\cellcolor{yellow}&\cellcolor{green}&&&\\
&&$N_1$&$N_1$&\cellcolor{blue}&\cellcolor{magenta}&\\
&&&&$N_2$&$N_2$&\cellcolor{black}\\
&&&&&&$N_3$
\end{tabular}

(b) WRDIFF, $\vec{\beta} = \{0, 1, 2 \}$
\end{subfigure}
\begin{subfigure}{13em}
\begin{tabular}{C{1.2em}C{1.2em}C{1.2em}C{1.2em}C{1.2em}C{1.2em}C{1.2em}}
$z_0$ & $z^*_1$ & $z_1$& $z^*_2$ & $z_2$& $z^*_3$ & $z_3$ \\
\cellcolor{red}&\cellcolor{orange}&&&&&\\
$N_0$&$N_0$&\cellcolor{yellow}&\cellcolor{green}&&\cellcolor{magenta}&\\
&&$N_1$&$N_1$&\cellcolor{blue}&$N_1$&\\
&&&&$N_2$&&\cellcolor{black}\\
&&&&&&$N_3$
\end{tabular}

(c) GRD, $\vec{\beta} = \{0, 1, 1 \}$
\end{subfigure}
\caption[]{Generalized recursive difference sample allocations for four models\footnotemark[4].  Labels below the column indicate the number of samples.  The arrows indicate that the $z_{i*}$ samples will move based on the values of $\vec{\beta}$.}
\label{fig:grd_diagram}
\end{figure}

In the same vein as the GMF estimators, the recursive difference sample allocation strategy associated with MLMC and WRDIFF can be generalized for any recursion structure.  In the recursive difference sample allocation strategy, the subsets $z_i$ of $z$ are disjoint (\ie independent). The elements of $\mathbf{G}$ and $\mathbf{g}$ for the generalized recursive difference (GRD) estimator are:
\begin{equation}
\begin{split}
G_{ij}^{GRD} =&  \left(\begin{array}{rl} \frac{1}{N_{\beta_i}} &: \beta_i=\beta_j\\ 0&: \beta_i \neq \beta_j\end{array}\right) 
- \left(\begin{array}{rl} \frac{1}{N_{\beta_i}} &: \beta_i=j\\ 0&: \beta_i \neq j\end{array}\right) 
- \left(\begin{array}{rl} \frac{1}{N_i} &: i=\beta_j\\ 0&: i \neq \beta_j\end{array}\right) 
+ \left(\begin{array}{rl} \frac{1}{N_i} &: i=j\\ 0&: i \neq j\end{array}\right) 
\end{split}
\end{equation}
\begin{equation}
\begin{split}
g_{i}^{GRD} =& \left(\begin{array}{rl} \frac{1}{N_0} &: \beta_i=0\\ 0&: \beta_i \neq 0\end{array}\right)
\end{split}
\end{equation}
The number of evaluations of model $i$ with GRD are $N_{i^* \cup i} = N_{\beta_i} + N_i$.  As with the GMF estimator the sub-optimization problem simplifies to optimizations of $\vec{N}$ given $\vec{\beta}$.
Schematics of sample allocations for the GRD estimator are shown in \fref{fig:grd_diagram} in addition to two examples of the estimator with given $\beta$.  The WRDIFF estimator is an instance of the GRD estimator with $\vec{\beta} = \{0, 1,\dots, M-1\}$.

\subsubsection{Generalized Independent Samples (GIS) Estimators}
\begin{figure}
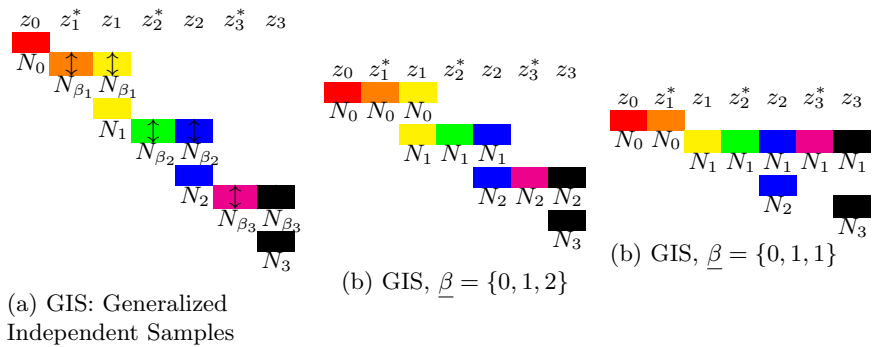

\centering
\small
\setlength\tabcolsep{1.5pt}
\renewcommand{\arraystretch}{0.7}

\begin{subfigure}{12em}
  \centering
\begin{tabular}{C{1.2em}C{1.5em}C{1.2em}C{1.5em}C{1.2em}C{1.5em}C{1.2em}}
$z_0$ & $z^*_1$ & $z_1$& $z^*_2$ & $z_2$& $z^*_3$ & $z_3$ \\
\cellcolor{red}&&&&&&\\
$N_0$&\cellcolor{orange} $\updownarrow$&\cellcolor{yellow} $\updownarrow$&&&&\\
&$N_{\beta_1}$&$N_{\beta_1}$&&&&\\
&&\cellcolor{yellow}&&&&\\
&&$N_1$&\cellcolor{green}$\updownarrow$&\cellcolor{blue}$\updownarrow$&&\\
&&&$N_{\beta_2}$&$N_{\beta_2}$&&\\
&&&&\cellcolor{blue}&&\\
&&&&$N_2$&\cellcolor{magenta}$\updownarrow$&\cellcolor{black}$\updownarrow$\\
&&&&&$N_{\beta_3}$&$N_{\beta_3}$\\
&&&&&&\cellcolor{black}\\
&&&&&&$N_3$
\end{tabular}
\caption{\raggedright GIS: Generalized \linebreak Independent Samples}
\end{subfigure}
\begin{subfigure}{12em}
  \centering
\begin{tabular}{C{1.2em}C{1.2em}C{1.2em}C{1.2em}C{1.2em}C{1.2em}C{1.2em}}
$z_0$ & $z^*_1$ & $z_1$& $z^*_2$ & $z_2$& $z^*_3$ & $z_3$ \\
\cellcolor{red}&\cellcolor{orange}&\cellcolor{yellow}&&&&\\
$N_0$&$N_0$&$N_0$&&&&\\
&&\cellcolor{yellow}&\cellcolor{green}&\cellcolor{blue}&&\\
&&$N_1$&$N_1$&$N_1$&&\\
&&&&\cellcolor{blue}&\cellcolor{magenta}&\cellcolor{black}\\
&&&&$N_2$&$N_2$&$N_2$\\
&&&&&&\cellcolor{black}\\
&&&&&&$N_3$
\end{tabular}
\caption{GIS, $\vec{\beta} = \{0, 1, 2 \}$}
\end{subfigure}
\begin{subfigure}{12em}
\begin{tabular}{C{1.2em}C{1.2em}C{1.2em}C{1.2em}C{1.2em}C{1.2em}C{1.2em}}
$z_0$ & $z^*_1$ & $z_1$& $z^*_2$ & $z_2$& $z^*_3$ & $z_3$ \\
\cellcolor{red}&\cellcolor{orange}&&&&&\\
$N_0$&$N_0$&\cellcolor{yellow}&\cellcolor{green}&\cellcolor{blue}&\cellcolor{magenta}&\cellcolor{black}\\
&&$N_1$&$N_1$&$N_1$&$N_1$&$N_1$\\
&&&&\cellcolor{blue}&&\\
&&&&$N_2$&&\cellcolor{black}\\
&&&&&&$N_3$
\end{tabular}

(b) GIS, $\vec{\beta} = \{0, 1, 1 \}$
\end{subfigure}
\caption[]{Generalized recursive difference sample allocations for four models \footnotemark[4].  Labels below the column indicate the number of samples. The arrows indicate that the $z_{i*}$ samples will move based on the values of $\vec{\beta}$.}
\label{fig:gis_diagram}
\end{figure}

It is also possible to generalize the independent sample strategy associated with ACVIS, which is very similar to GRD.  The independent samples strategy deviates form the recursive difference strategy in that the $z_{i*}$ samples are added to $z_i$.  Thus for GIS, $z^*_i \subset z_i$ and $\beta_i = j$ implies $z^*_i = z'_j$ where $z'_j = z_j - z^*_j$.  The elements of $\mathbf{G}$ and $\mathbf{g}$ for the generalized independent samples (GIS) estimator are:
\begin{equation}
\begin{split}
G_{ij}^{GIS} =&  \left(\begin{array}{rl} \frac{1}{N_{\beta_i}} - \frac{1}{N_{\beta_i} + N_i} - \frac{1}{N_{\beta_j} + N_j} + \frac{N_{\beta_i}}{(N_{\beta_i} + N_i)(N_{\beta_j} + N_j)}&: \beta_i=\beta_j\\ 0&: \beta_i \neq \beta_j\end{array}\right) \\
&+  \left(\begin{array}{rl}\frac{N_{\beta_i}}{(N_{\beta_i} + N_i)(N_{\beta_j} + N_j)} - \frac{1}{N_{\beta_j} + N_j} &: \beta_i=j\\ 0&: \beta_i \neq j \end{array}\right) \\
&+  \left(\begin{array}{rl}\frac{N_i}{(N_{\beta_i} + N_i)(N_{\beta_j} + N_j)} - \frac{1}{N_{\beta_i} + N_i} &: \beta_j=i\\ 0&: \beta_j \neq i \end{array}\right) 
+  \left(\begin{array}{rl}\frac{N_i}{(N_{\beta_i} + N_i)(N_{\beta_j} + N_j)} &: i=j\\ 0&: i \neq j \end{array}\right) \\
\end{split}
\end{equation}
\begin{equation}
\begin{split}
g_{i}^{GRD} =& \left(\begin{array}{rl} \frac{1}{N_0} - \frac{1}{N_0 + N_i}&: \beta_i=0\\ 0&: \beta_i \neq 0\end{array}\right)
\end{split}
\end{equation}
The number of evaluations of model $i$ with GIS are $N_{i^* \cup i} = N_{\beta_i} + N_i$.  Note that in GIS the values of $\vec{N}$ do not represent the size of $z_i$, but rather they represent the size of $z'_i$. As with the GMF and GRD estimators the sub-optimization problem simplifies to optimizations of $\vec{N}$ given $\vec{\beta}$.
Schematics of sample allocations for the GIS estimator are shown in \fref{fig:gis_diagram}, in addition to two examples of the estimator with given $\beta$.

\subsection{Relaxation of Ordering Constraints}
In most existing ACV estimators, model order affects variance reduction. For instance, the MFMC estimator assumes $N_i < N_{i+1}$. The ACVMF estimator optionally enforced this same ordering, with the authors noting it led to more robust numerical optimization \cite{gorodetsky_2018}.  Even in the unordered case, the ACVMF estimator retained the constraint that $N_0 < N_i$.\footnote{The ACVMF implementation in the current work enforces the ordering constraint.}

When model costs and covariance are similar between low-fidelity models, the optimal model ordering may not be apparent. This situation can arise when multiple machine learning models are used without a clear ``best.'' Notably, the GMF and GRD estimators make no assumption on ordering, except that $i=0$ represents the high-fidelity model. Enforcement of ordering constraints was found to have a negligible impact on optimization robustness; thus, no ordering constraints are included in these models whatsoever (\ie even the $N_0 < N_i$ constraint is relaxed).

To illustrate the effect of relaxed ordering constraints, a new estimator dubbed ACVMFU is constructed.  ACVMFU is the GMF estimator with $\beta=\{0, \dots, M-1\}$ and is equivalent to an ACVMF estimator with no ordering constraints.  Comparisons of ACVMF and ACVMFU estimators are presented in the results sections.

\subsection{Automatic Model Selection}
The choice of which models to include in an ACV estimator has an effect on the resulting accuracy of the estimator.  This is primarily due to the fact that estimators assume that all models must have samples allocated to them.  Since relaxing this constraint is not always practical within a sub-optimization, an alternative approach is to perform multiple sub-optimizations with differing subsets of models.  In this way sub-optimizations are parametrically defined based on subsets of $\{Q_0, Q_1, ..., Q_M\}$; thus, allowing for the number of samples allocated to a specific model to be 0 in the case that the model is not included in the sub-optimization.  Importantly, the high fidelity model must be a part of all model subsets in order for the estimators to remain unbiased.

Note that this type of parametrically defined sub-optimization is relatively straight forward and a form of it may already be used by many in practice. However, current practice usually relies on human intuition in deciding a limited number of model subsets to investigate.  As will be shown in later sections, the optimal subsets are not always intuitive; thus, automatic model selection which investigates all possible subsets of the models is preferred to manual subset selection.

A useful consequence of parametrically defining sub-optimizations based on model subsets is that a comparison is automatically made to the standard MC estimator (obtained when the model subset is  $\{Q_0\}$).  This comparison is standard for the multi-model MC community.

%% file: algorithm_summary.tex
\section{Algorithm Summary}
\label{sec:algo_summary}
The 13 ACV algorithms discussed in proceeding sections are summarized in Table \ref{tab:algorithm_summary}.  The algorithms are described in terms of optimization type, sampling strategy, and recursion structure.  In many cases, an algorithm is completely dominated by others.  In this context, algorithm domination means that the optimization domain of one algorithm is contained within the optimization domain of the dominating algorithm.  The dominating algorithm cannot perform worse than the dominated one if identical, deterministic (sub)optimizations are used.  The domination organization is described by: 
$$
\text{GRDMR} \leq \text{GRDSR} \leq \text{WRDIFF} \leq \text{MLMC} \footnote{In some cases MLMC may perform better than its dominating algorithms due to the difference in analytical versus numerical optimization.}\\
$$
$$
\text{GISMR} \leq \text{GISSR} \leq \text{ACVIS}\\
$$
$$
\text{ACVKL} \leq \text{MFMC}\\
$$
$$
\text{ACVKL} \leq \text{ACVMF}\\
$$
$$
\text{GMFMR} \leq \text{GMFSR} \leq \text{ACVMFU} \leq \text{MFMC}\footnote{In some cases MFMC may perform better than its dominating algorithms due to the difference in analytical versus numerical optimization.}
$$
where the notation A $\leq$ B is used to indicate that algorithm A dominates algorithm B.

\begin{table}
\centering
\caption{Algorithm Summary}
\label{tab:algorithm_summary}
\begin{tabular}{|c|c|c|c|c| }
\hline
Algorithm & Optimization & $z_i$ & $\vec{\beta}$ &Notes \\ \hline\hline

\multicolumn{5}{|c|}{\bf Recursive Difference Algorithms} \\ \hline
MLMC & analytic &$z_i \cap z_j = \emptyset$ for $i \neq j$& $\beta_i = i-1$ &   $\alpha_i = -1$ \cite{Giles08multi-levelmonte, 10.1007/978-3-642-41095-6_4} \\ \hline
WRDIFF & numerical &$z_i \cap z_j = \emptyset$ for $i \neq j$& $\beta_i = i-1$ & \cite{gorodetsky_2018} \\ \hline
GRDSR & numerical & $z_i \cap z_j = \emptyset$ for $i \neq j$& $\vec{\beta}^{SR}$ & \\ \hline
GRDMR & numerical & $z_i \cap z_j = \emptyset$ for $i \neq j$& $\vec{\beta}^{MR}$ & \\ \hline

\multicolumn{5}{|c|}{\bf Independent Samples Algorithms} \\ \hline
ACVIS & numerical &$z_i \cap z_j = z_0$ for $i \neq j$& $\beta_i = 0$ & \cite{gorodetsky_2018} \\ \hline
GISSR & numerical &\begin{tabular}{@{}c@{}}$z_i = z_i^* \cup z_i'$ for $i > 0$, \\  $z_i' \cap z_j' = \emptyset$ for $i \neq j$\end{tabular} & $\vec{\beta}^{SR}$  &$\beta_i =j \rightarrow z_i^* = z_j'$ \\ \hline
GISMR & numerical &\begin{tabular}{@{}c@{}}$z_i = z_i^* \cup z_i'$ for $i > 0$, \\  $z_i' \cap z_j' = \emptyset$ for $i \neq j$\end{tabular} & $\vec{\beta}^{MR}$  &$\beta_i =j \rightarrow z_i^* = z_j'$ \\ \hline

\multicolumn{5}{|c|}{\bf Multifidelity Algorithms} \\ \hline
MFMC & analytic & $z_{i-1} \subset z_i$ for $i > 0$& $\beta_i = i-1$ &$N_i < N_{i+1}$ \cite{mfmc2016} \\ \hline
ACVMF & numerical &$z_{i-1} \subset z_i$ for $i > 0$& $\beta_i = 0$ &$N_i < N_{i+1}$ \cite{gorodetsky_2018} \\ \hline
ACVKL & numerical & $z_{i-1} \subset z_i$ for $i > 0$& $\beta^{KL}$ &   $N_i < N_{i+1}$ \cite{gorodetsky_2018}\\ \hline
ACVMFU & numerical &  $z_i \subset z_j$ or $z_j \subset z_i$& $\beta_i = 0$ & \\ \hline
GMFSR & numerical &  $z_i \subset z_j$ or $z_j \subset z_i$& $\vec{\beta}^{SR}$  & \\ \hline
GMFMR & numerical &  $z_i \subset z_j$ or $z_j \subset z_i$& $\vec{\beta}^{MR}$  & \\ \hline
\end{tabular}

\end{table}

%% file: practical_considerations.tex
\section{Practical Considerations}\label{sec:practical}
This section discusses details that are hopefully useful when implementing ACV estimators in pactice. Because most of the ACV estimators discussed in this work require numerical optimization, the majority of this section is devoted to improving robustness and accuracy of that task.

\subsection{Real-valued Sample Numbers}
Following previous works \cite{mfmc2016,gorodetsky_2018}, restriction of sample numbers to integer values (\ie $N_i \in \mathbb{N}$) is relaxed to allow for the use of real-valued optimization, resulting in unrealistic sample numbers $N_i \in \mathbb{R}$. After each sub-optimization the floor $\left \lfloor{N_i}\right \rfloor $ is taken and variance is recalculated; this ensures the sub-optimization result satisfies the total cost constraint and that the variance is accurate. The effect of flooring on variance is assumed to be small (\ie does not greatly alter the sub-optimization landscape). A constraint of $N_i >1$ is used to avoid flooring to $N_i = 0$. To avoid issues when $z^*_i = z_i$, constraints on the optimization parameters ensure $z^*_i \neq z_i$ after flooring.  For example, the GMF estimator constrains $|N_{\beta_i} - N_i| > 1$. GRD and GIS estimators do not need such a constraint.

\subsection{Total Cost Constraint}
The tunable recursion estimators (GMF, GRD and GIS), have $M + 1$ optimization parameters associated with $\vec{N} = \{N_0, \dots, N_M\}$.  As was done with the ACVMF, ACVIS, and ACVKL estimators in \cite{gorodetsky_2018}, the total cost constraint can be handled implicitly by defining ratios $r_i = N_i / N_0$, where $i \in \{1, \dots, M\}$.  The value of $N_0$, and subsequently $N_i$, can be calculated from the total cost constraint and model costs:
\begin{equation}
N_0 = \frac{\tilde w ^*}{ w_0 + \sum\limits_{i=1}^M w_ir_{i^* \cup i}}
\end{equation}
where $r_{i^* \cup i}$ is the similarly normalized number of model $i$ evaluations; \eg for the GRD estimator $r_{i^* \cup i} = r_{\beta_i} + r_i$.  This handling of the total cost constraint has the desirable effect of reducing the optimization space from $M + 1$ parameters to $M$ parameters. 
Note that the total cost of an optimized estimator will be less than the target cost due to rounding.  No effort is made here to systematically add model evaluations or ``top off'' the estimator after rounding.

\subsection{Numerical Optimization Details}
This subsection details the two-stage, black-box, numerical optimization approach used herein. The authors make no claim that this numerical optimization scheme is ideal, but rather that it was effective in performing the required ACV estimator sub-optimizations.  Both stages utilize the Python package SciPy \cite{scipy}.  In the first stage, a gradient-based optimization is performed using the SLSQP algorithm and previously described constraints.  The second stage is a gradient-free simplex optimization using the Nelder-Mead algorithm. A two stage approach was chosen because the gradient based method is able to quickly find the general region of the minimum but has trouble pinpointing the minimum itself. Thus, optimization time is reduced relative to using the simplex method only.

All optimizations are initialized with $r_i = i +1$ for $i \in \{1, \dots, M\}$  such that all constraints are satisfied. Smarter starting points could be chosen for each optimization but were fixed here to control for associated effects. From the starting point, SLSQP optimization is performed.  The objective function was the variance of the ACV estimator. The gradient was calculated with automatic differentiation using PyTorch \cite{pytorch}, and the SLSQP function tolerance was set to $10^{-10}$. The SLSQP optimization result serves as the starting point for the second stage. Occasionally, SLSQP has difficulty satisfying the specified constraints and results in an infeasible optimization result. In this case, the original fixed starting point is used for the second stage.

The second stage is a Nelder-Mead simplex optimization where constraints are enforced using a penalization scheme. Each constraint $C_i > 0$ is enforced by a penalty
\begin{equation}
P_i = \left\{ \begin{array}{rl} -10^{16}C_i& : C_i<0\\ 0&: C_i\geq0\end{array}\right.
\end{equation}
applied to the objective function (\ie the ACV estimator variance). In this stage, function evaluations were limited to $500\times M$ and an absolute tolerance of $10^{-12}$ on the parameters and objective function was used.



%% file: literature_examples.tex
\section{Literature Examples}\label{sec:lit_examples}
Two examples from the work of Gorodetsky et al. \cite{gorodetsky_2018} are revisited here, to illustrate the performance of the proposed parametrically-defined ACV estimators relative to existing estimators.

\subsection{Monomial Example}
The first example is defined by a set monomial models, $Q_i(z) = z^{5-i}$ for $i \in \{0, \dots, 4\}$ and $z \sim \mathcal{U}(0, 1)$.  The covariance matrix for these models is defined analytically as $\Cov[Q_i, Q_j] = \frac{1}{11 - i - j} - \frac{1}{(6-i)(6-j)}$.  The model costs, $\vec{\omega}$, are chosen such that each model is 10 times cheaper than the previous model: $\omega_i = 10^{-i}$ for $i \in \{0, \dots, 4\}$.  Additionally, the effect of model costs on variance reduction and algorithm performance is investigated using a second set of model costs.  In this scenario all of the low-fidelity models are 10 times faster than in the first scenario: $\omega_0 = 1$ and $\omega_i = 10^{-i-1}$ for $i \in \{1, \dots, 4\}$.  These two model cost scenarios are henceforth referred to as no-cost-gap and cost-gap, respectively.  In both scenarios, a target cost of 20 is used.

\begin{figure}
\begin{subfigure}[b]{.4\textwidth}
  \includegraphics[scale=1.0]{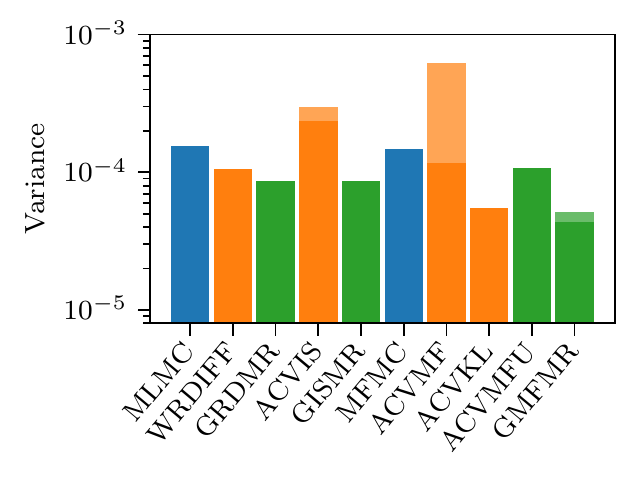}
  \caption{No-cost-gap Scenario}
\end{subfigure}
\begin{subfigure}[b]{.4\textwidth}
  \includegraphics[scale=1.0]{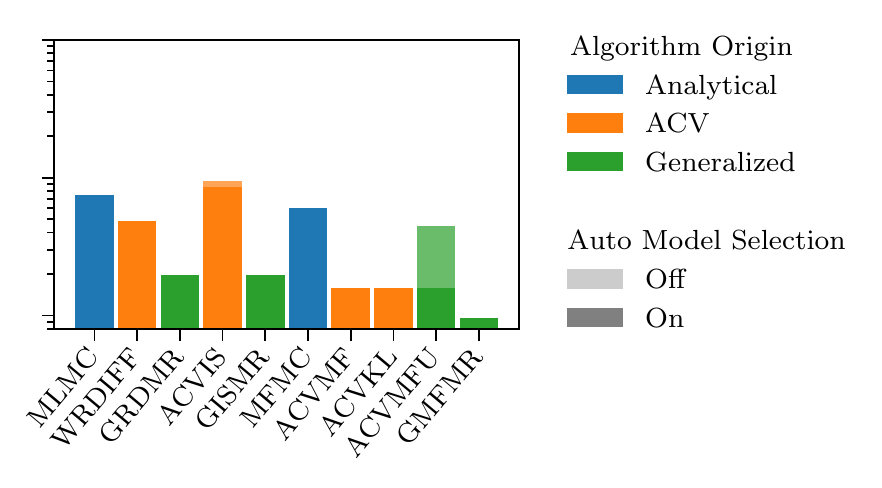}
  \caption{Cost-gap Scenario}
\end{subfigure}
\caption{Variance of ACV estimators for the monomial example for 10 algorithms.  The darker regions of each bar indicate automatic model section is on. The lighter region of each bar indicates automatic model selection is off.}
\label{fig:monomial_performance}
\end{figure}

The variance of the ACV estimator for $\E[Q_0]$ of the monimial example was found for 10 of the algorithms defined in \sref{sec:algo_summary}; the results are illustrated in \fref{fig:monomial_performance}.  Each algorithm was used both with and without automatic model selection.  In both model cost scenarios, the GMFMR algorithm provides the estimator with lowest variance.  More generally, the parametrically-defined estimators had better variance reduction than their more restricted counterparts. 

The effect of ordering constraints (seen by comparing ACVMF to ACVMFU) is inconclusive for this example, since ACVMFU performed better in one scenario and ACVMF in the other.  Additionally, it appears that automatic model selection may reduce the influence of the constraints since the results for ACVMF and ACVMFU are equivalent when it is being used.

\begin{figure}
\begin{subfigure}[b]{.4\textwidth}
\hspace{8pt}
\includegraphics[scale=1.0]{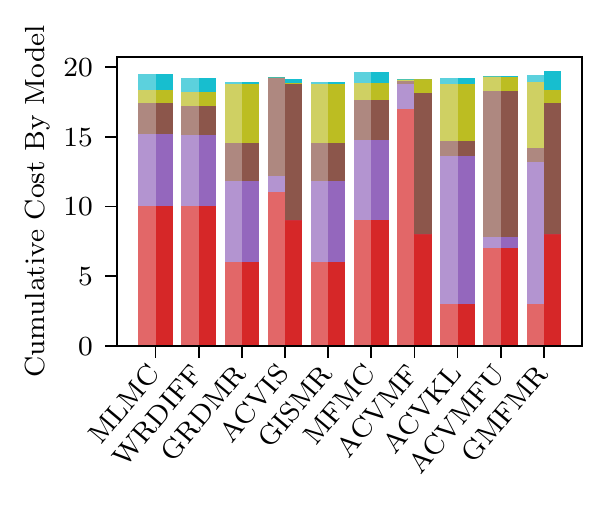}  \caption{No-cost-gap Scenario}
\end{subfigure}
\begin{subfigure}[b]{.4\textwidth}
\includegraphics[scale=1.0]{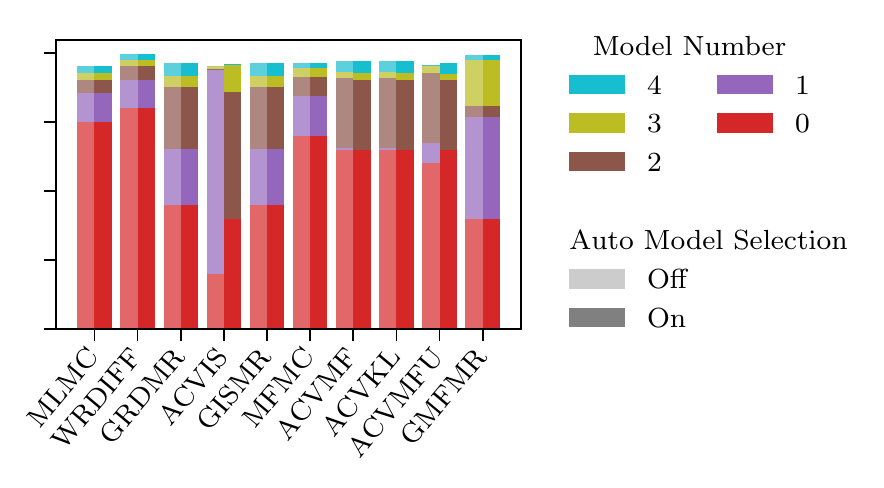}
  \caption{Cost-gap Scenario }
\end{subfigure}
\caption{Allocation of cost to different models in the monomial example.  The darker regions of each bar indicate automatic model section is on. The lighter region of each bar indicates automatic model selection is off.}
\label{fig:monomial_allocation}
\end{figure}

The effect of model costs is seen by comparing the model cost scenarios in \fref{fig:monomial_performance} (a) and (b).  Naturally, reducing the cost of the low-fidelity models leads to better variance reduction for each algorithm.  However, the rankings of the algorithms is relatively invariant to the change in model costs.  The effect of model cost on the relative performance of the algorithms is greater with the multifidelity algorithms than the independent samples or recursive difference algorithms.  The amount of the total cost attributed to each of the models (for each algorithm) is shown in \fref{fig:monomial_allocation}.  Interestingly, most algorithms tend to place increased emphasis on the high-fidelity model in the cost-gap scenario rather than on the low-fidelity models.  Thus it seems that the majority of variance reduction in the cost-gap scenario stems from the ability to afford more high-fidelity model samples.

Automatic model selection was not very beneficial for most algorithms in the monomial example, potentially because all models are valuable.  Note, however, that the best allocation for the no-cost-gap scenario is found with automatic model selection.  Looking at the allocations in \fref{fig:monomial_allocation} can give further insight into automatic model selection.  Automatic model selection led to better allocations in 3 no-cost-gap algorithms: in all 3 it identified Model 1 for removal (the highest cost low-fidelity model), effectively allowing for greater emphasis on the faster models.  In ACVMF, ACVKL, and ACVMFU  for the cost-gap scenario automatic model selection removes Model 1, but ultimately leads to very similar allocations (and very similar variance).  Other examples do, however, illustrate that automatic model selection can lead to drastically different allocations (\eg ACVIS in the cost-gap scenario).

In both model cost scenarios, GMFMR with automatic model selection (the best algorithm) produces drastically different allocations than the other algorithms.  This illustrates that the extension of the ACV variance reduction optimization to parametrically-defined subdomains does result in better variance reduction.

\subsection{Wave Propagation Example}
The second example focusses on elastic wave propagation in heterogeneous media in two spatial dimensions.  The example is briefly described here.  For a fully-detailed description of the example refer to the work of Gorodetsky et al. \cite[Section 4.3]{gorodetsky_2018}.  

The wave propagation example describes the propagation of a shockwave in a bi-material 2-D domain, where the material properties of the sub-domains are uncertain. Three scenarios are considered in this example; the same high fidelity model is used for all scenarios.  In the first scenario, dubbed the multilevel scenario, four low-fidelity models are used, derived from spatial coarsening of the high-fidelity model. Each of the models, $Q_i$, have roughly a fourth of the computational cells of the previous model.  In the second scenario, dubbed the multifidelity scenario,  the four low-fidelity models have the same spatial resolution as the multi-level scenario; however, a lower order discretization is used.  The third scenario, dubbed the multifidelity with gap scenario, is the same as the multi-fidelity scenario but with the finest, low-fidelity model ($Q_1$) removed.  A target (normalized) cost of 30 was used for all scenarios.

\begin{figure}
\centering
\begin{subfigure}{.44\textwidth}
\includegraphics[scale=1]{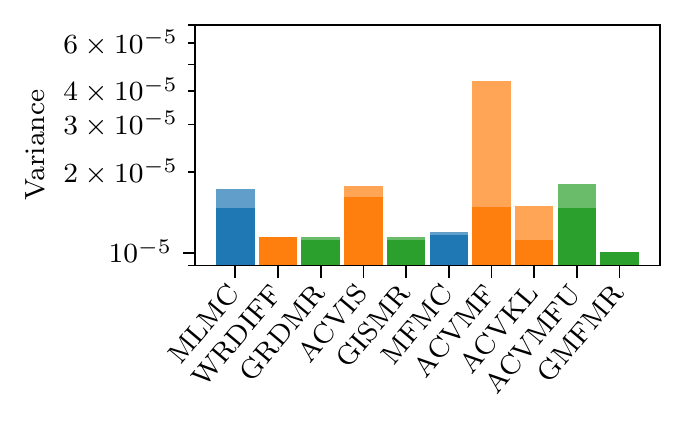}
  \caption{Multi-level Scenario}
\end{subfigure}
\begin{subfigure}{.4\textwidth}
\includegraphics[scale=1]{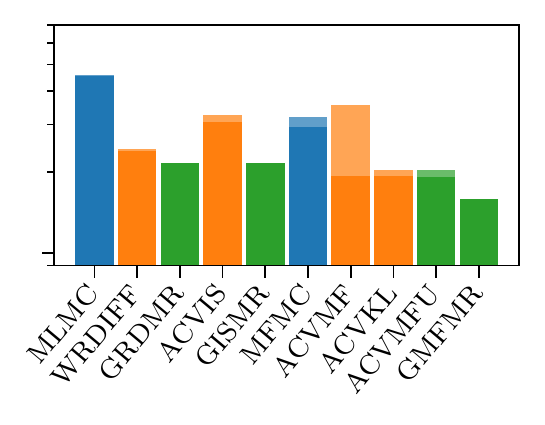}
  \caption{Multifidelity Scenario}
\end{subfigure}
\begin{subfigure}{.6\textwidth}
\includegraphics[scale=1]{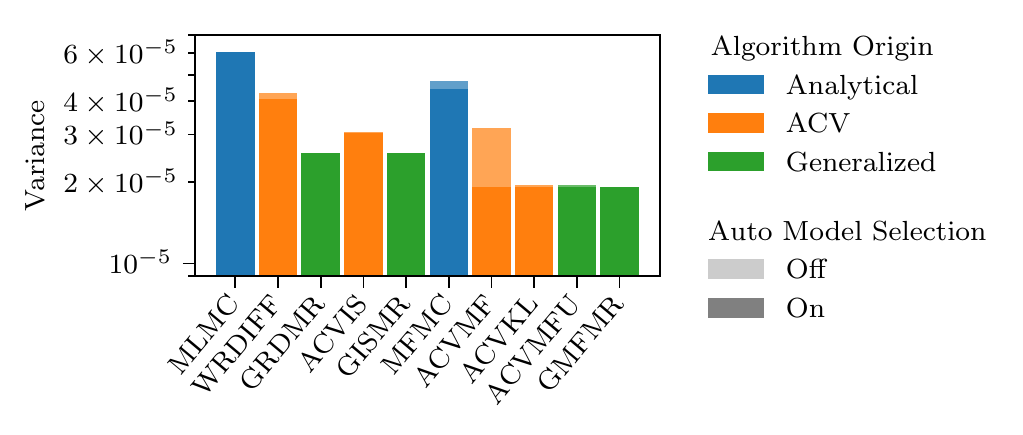}
  \caption{Multifidelity-with-gap Scenario}
\end{subfigure}
\caption{Variance of ACV estimators for the wave propagation example. The darker regions of each bar indicate automatic model section is on. The lighter region of each bar indicates automatic model selection is off.}
\label{fig:wave_prop_variance}
\end{figure}

Again, 10 algorithms were run for each scenario, both with and without automatic model selection.  The variances in the resulting estimators are illustrated in \fref{fig:wave_prop_variance}.  The multi-level scenario generally led to better variance reduction than the multifidelity scenarios.  However, with the exception of MFMC,  the ranking of the algorithms is relatively invariant to the scenarios.
In the multi-level scenario, MFMC performs better relative to ACVMF and ACVMFU.  This may be because the recursion structure allows it to better take advantage of all the models in a nested recursion structure.  Note that ACVKL and GMFMR, which include the nested recursion structure of MFMC in their suboptimizations, still outperform MFMC.
Most algorithms perform worse in the multifidelity-with-gap scenario than the multifidelity scenario due to the exclusion of a model.  ACVIS, ACVMF, ACVKL, and ACVMFU perform about equally well before and after the model exclusion.

In this example, since ACVMFU always outperforms ACVMF, the relaxation of ordering constraints was advantageous.  However, as in the monimial example, the use of automatic model selection diminishes the effect of the constraints.

\begin{figure}
\centering
\begin{subfigure}{1.0\textwidth}
\centering
\includegraphics[scale=0.47]{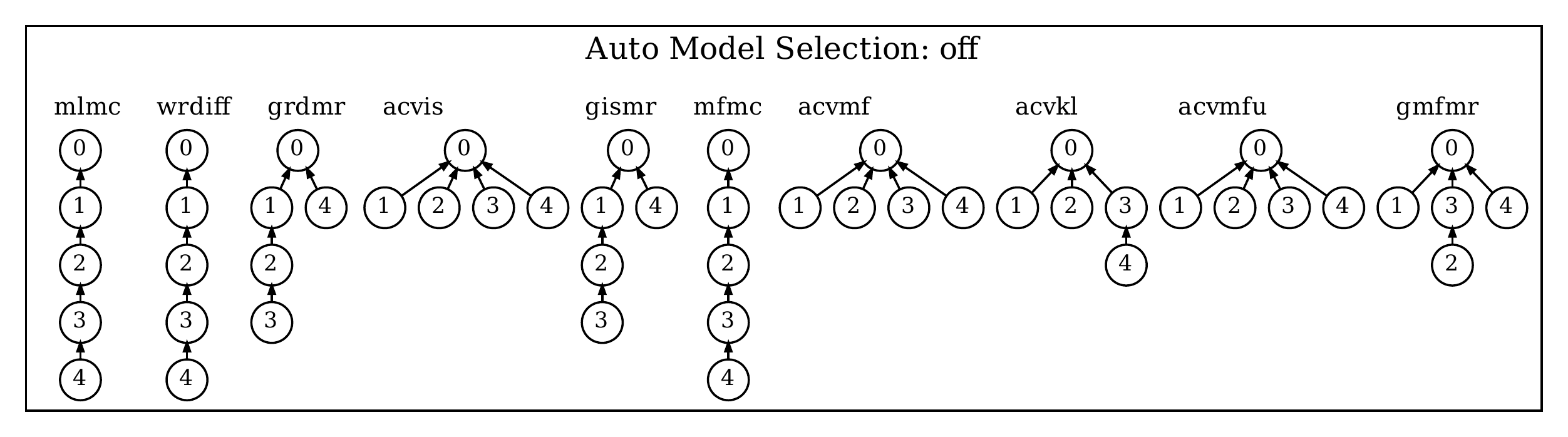}
\includegraphics[scale=0.47]{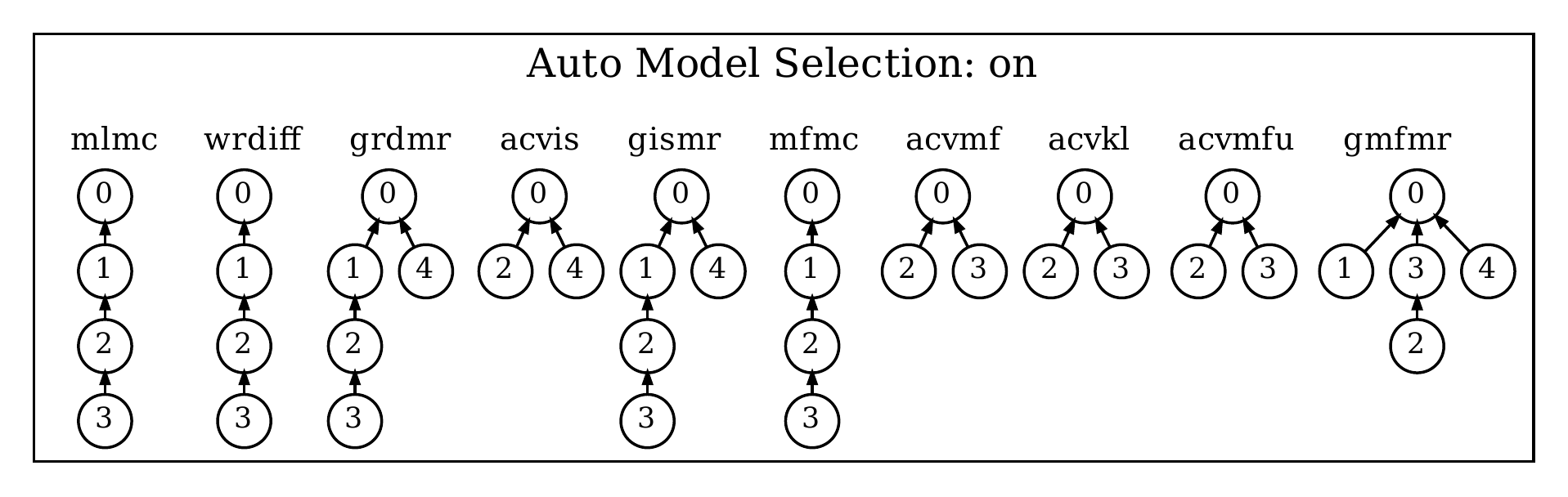}
  \caption{Multifidelity Scenario}
\end{subfigure}
\begin{subfigure}{1.0\textwidth}
\centering
\includegraphics[scale=0.47]{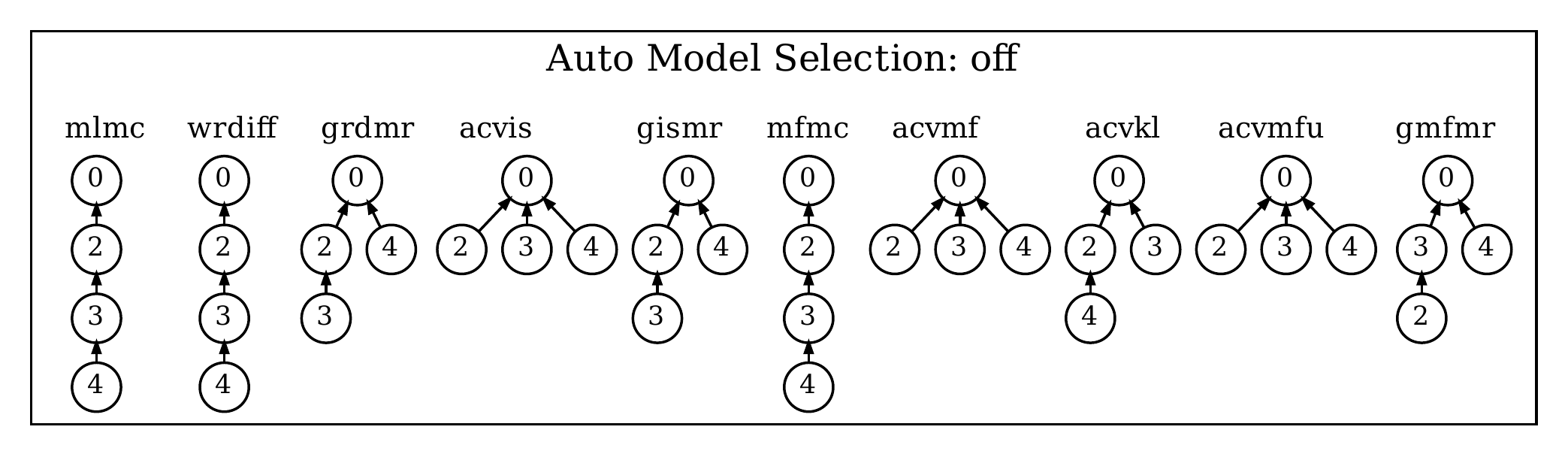}
\includegraphics[scale=0.47]{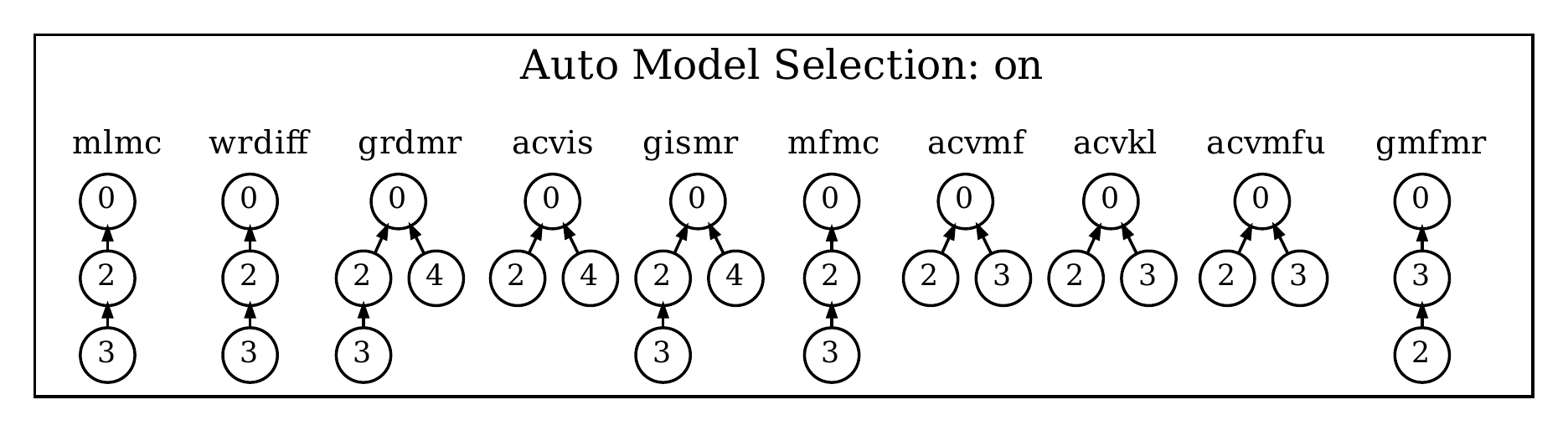}
  \caption{Multifidelity-with-gap Scenario (Model 1 removed)}
\end{subfigure}
\caption{Recursion structures of the estimators produced for the multifidelity wave propagation example.  $\raisebox{.5pt}{\textcircled{\raisebox{-.4pt} {\scriptsize $i$}}}\leftarrow$ \raisebox{.5pt}{\textcircled{\raisebox{-.4pt} {\scriptsize $j$}}} indicates that model $j$ acts as a control variate for model $i$.}
\label{fig:wave_prop_recursion}
\end{figure}

Automatic model selection was more useful in this example than the monimial example; however, it often excluded models with only a minor effect on the variance reduction. In these cases, the number of samples asscoiated with the model were low.  For instance, in both of the multifidelity scenarios, MLMC excludes Model 4 without an appreciable difference in variance.   The recursion structures of the estimators for the multifidelity scenarios are illustrated in \fref{fig:wave_prop_recursion}.  These recursion structures illustrate the effect that model exclusion and automatic model selection can have on the structure of the estimators.  Most apparent from this visualization, is that the use of automatic model selection results in the use of subsets of models that are not always intuitive.  For instance ACVIS, ACVMF, ACVKL, and ACVMFU  all identify that Model 1 and 4 aren't very useful.  It seems unlikely that a user would be able to identify from the covariance matrix / model costs that these models should be excluded to achieve best variance reduction for these methods.  This illustrates the value of parametric subdomain generation, such that a researcher need not manually try all combinations of model subsets.
Note that the above algorithms, which automatically removed Model 1 and 4, are the ones least effected by the exclusion of Model 1 in the multifidelity-with-gap scenario.

In all scenarios, GMFMR resulted in best variance reduction.  Looking again at \fref{fig:wave_prop_recursion}, GMFMR resulted in a recursion structure that is not intuitive, and is unique from any previously published ACV method.  Interestingly, a finer model acts as a control variate for a coarser one:  Model 2 acts as a control variate for Model 3.  This again shows the value of parametric generation of subdomains for optimization.

%% file: parametric_study.tex
\section{Randomly Selected Models Example}\label{sec:parameter_sweep}
The literature examples of \sref{sec:lit_examples} are not representative of all possible model scenarios. To study broader algorithm performance, a procedure for generating random, admissible model scenarios was developed.  A model scenario can be parameterized by the elements of the model covariance matrix, $\Cov\left[Q_i , Q_k \right]$, and the associated model costs, $\vec{w}$, both of which exhibit implicit dependence on the number of models ($M+1$). 

Random correlation matrices were sampled from the Lewandowski-Kurowicka-Joe (LKJ) distribution \cite{lkj_2009_gencorr}. This distribution can be used to uniformly sample the space of positive definite correlation matrices and, thus, is commonly used as a prior distribution for the covariance matrix of multivariate normal distributions. The LKJ distribution is parameterized by the number of models and shape parameter, $\eta$, which can be used to tune the correlation strength between elements. Specifically, $\eta<1$ favors stronger correlations (i.e., off-diagonal values are closer to one) and $\eta>1$ favors weaker correlations (i.e., off-diagonal values are closer to zero). For $\eta=1$ the density is uniform over correlation matrices of order equal to the number of models.

For each fixed number of models $\in\{2,3,4,5,6\}$, $1\times10^6$ model scenarios were generated using the LKJ distribution with $\eta=1$. Due to computational restrictions, $5\times10^5$ scenarios were generated with 5 models and $1.6\times10^5$ scenarios were generated with 6 models. Model variances were required to develop covariance matrices from the generated correlation matrices. The variance of the highest fidelity model was fixed such that $\Var[Q_0]=1$. The remaining variances for $i=1,\ldots,M$ were assumed to be independent and identically distributed (iid) random variables and $\Var[Q_i] \sim U(0.1, 1.5)$. Total cost was fixed $\tilde{w}^*=1$ and the high fidelity model cost $w_0=\tilde{w}^*/100$. The remaining elements of the cost vector were obtained by assuming iid ratios, $\log_{10}(w_i/w_0) \sim U(-6, 0)$. The algorithms of \sref{sec:background} and \sref{sec:parametric_estimators} were used to derive sample allocations minimizing estimator variance for each model scenario. Results from these optimizations are presented and discussed in following subsections.

The results are discussed in context of a metric called mean relative deviation, $\bar D$.  This metric was utilized in order to control for the fact that covariance and model costs in each model scenario highly influence variance reduction for all algorithms.  Mean relative deviation is defined as the average fractional increase of variance of an algorithm relative to the best algorithm:
\begin{equation}
\bar{D} = \frac{1}{N_{ms}} \sum_{i=1}^{N_{ms}} \frac{\Var[\tilde {Q}]^{(i)} - V^{(i)}_B}{V^{(i)}_{B}}
\end{equation}
$V^{(i)}_B$ is the variance of the best algorithm for model scenario $i$ among a set of algorithms, $\mathcal{D}_{algs}$.  In the remainder of this section, the algorithms in $\mathcal{D}_{algs}$ will vary; this allows for direct comparison of algorithms in smaller subsets. The $\mathcal{D}_{algs}$ used in evaluation of $\bar D$ is specified in each of the subsections to follow.

\subsection{Aggregate Algorithm Performance}
\begin{figure}
  \begin{subfigure}[T]{.425\textwidth}
\includegraphics[scale=1.0]{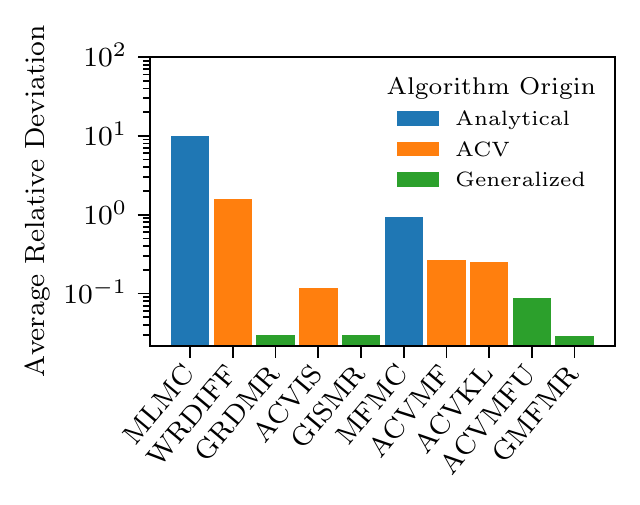}
  \caption{}
  \label{fig:aggregate_performance_all}
\end{subfigure}
  \begin{subfigure}[T]{.4\textwidth}
\includegraphics[scale=1.0]{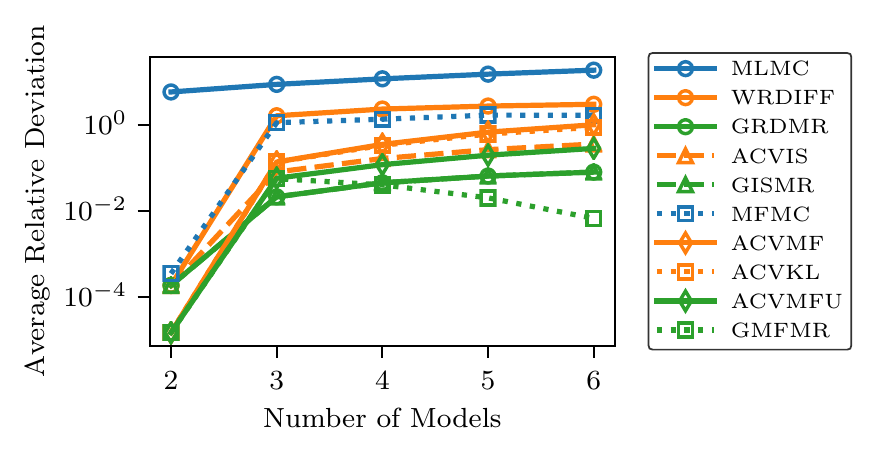}
  \vspace{1pt}
  \caption{}
  \label{fig:aggregate_performance_by_m}
\end{subfigure}
\caption{Aggregate algorithm performance (a) on all random model scenarios and (b) as a function of the number of models.}
\label{fig:aggregate_performance}
\end{figure}

In order to compare to the model scenarios of the previous section, the mean relative deviation was calculated for all random model scenarios and the same algorithms in \frefs{fig:monomial_performance}{fig:wave_prop_variance}: $\mathcal{D}_{algs} = \{$MLMC, WRDIFF, GRDMR, ACVIS, GISMR, MFMC, ACVMF, ACVKL, ACVMFU, GMFMR$\}$.  The result, which is illustrated in \fref{fig:aggregate_performance_all}, are similar to the literature examples in that the generalized methods outperform their more restrictive counterparts.  There are two differences here compared to the literature examples.  Firstly, GRDMR and GISMR perform slightly better than GMFMR, indicating that sampling strategies (\ie independent samples, recursive difference, and multifidelty) may be more/less ideal based on the model scenario.  Secondly, it is now clear that ACVMFU performs better than ACVMF, meaning that the relaxation of ordering constraints is generally beneficial.

The relationship between the number of models and mean relative deviation is illustrated in \fref{fig:aggregate_performance_by_m}.  For this figure, the mean deviation was calculated based on model scenarios grouped by the number of models.  As one might expect, the algorithms all perform similarly with two models ($M=1$) since the differences between each of the algorithms and sampling strategies is very limited.  The exception here is MLMC which performs significantly worse than the others because of its non-optimal $\alpha$.  Beyond two models  the algorithms begin to perform more distinctly.  For all algorithms except GMFMR the slope with $M \ge 2$ is positive.  This indicates that as the number of models increases, the GMFMR algorithm increasingly outperforms the others.

\subsection{Influence of Effective Optimization Domain Size on Performance}
The effect of increasing the effective optimization domain on variance reduction is investigated in this subsection.  To control for the effect of sampling strategy on variance reduction, the algorithms are broken into four categories as described in \tref{tab:alg_groupings}.  For each of these categories the algorithms are compared to each other (i.e. $\mathcal{D}_{algs}$ contains the same algorithms as the category).  In each of the categories there is a range of algorithms from most to least restricted optimization domain.  The model scenarios are grouped into sets based on the relative performance of the algorithms in that group to illustrate how a change in optimization domain corresponds to algorithm performance.

\begin{table}
\centering
\caption{Algorithms grouped by sampling strategy.}
\label{tab:alg_groupings}
\begin{tabular}{|c|c|}
\hline
Sampling Strategy & Algorithms ($\mathcal{D}_{algs}$)\\ \hline \hline
Independent Samples & ACVIS, GISSR, GISMR \\\hline
Recursive Difference & MLMC, WRDIFF, GRDSR, GRDMR\\\hline
Ordered Multifidelity & MFMC, ACVMF, ACVKL\\\hline
Unordered Multifidelity & ACVMFU, GMFSR, GMFMR\\\hline
\end{tabular}
\end{table}

The independent samples algorithms distinguish themselves from one another solely by the recursion trees searched in their suboptimizations. ACVIS searches a single recursion tree. GISSR searches all recursion trees of maximum depth 3; this includes the ACVIS recursion tree.  GISSMR searches all recursion trees; this includes all GISSR trees and the ACVIS tree.

\begin{figure}
\centering
\begin{subfigure}{0.45\textwidth}
\includegraphics[scale=1]{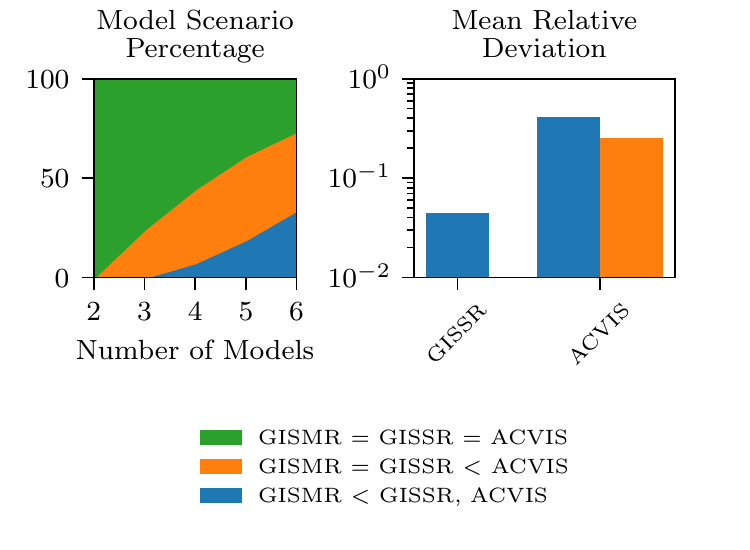}
\caption{Independent Samples}
\label{fig:independent_samples}
\end{subfigure}
\begin{subfigure}{0.45\textwidth}
\includegraphics[scale=1]{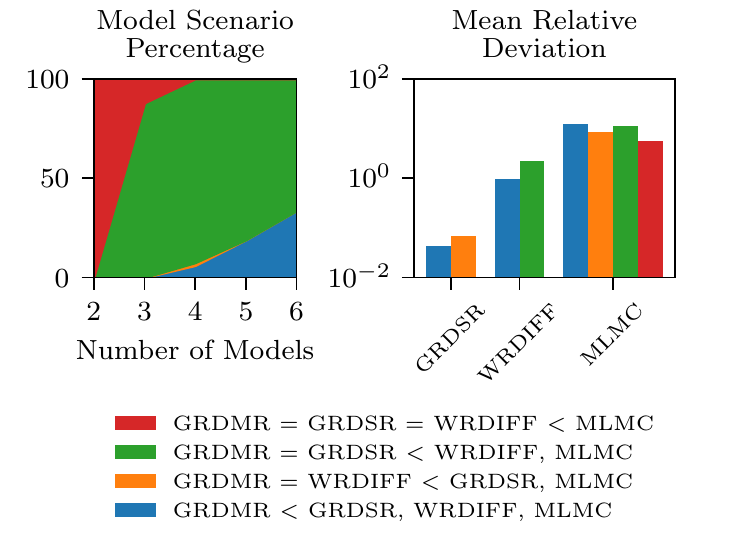}
\caption{Recursive Difference}
\end{subfigure}
\begin{subfigure}{0.45\textwidth}
\vspace{10pt}
\includegraphics[scale=1]{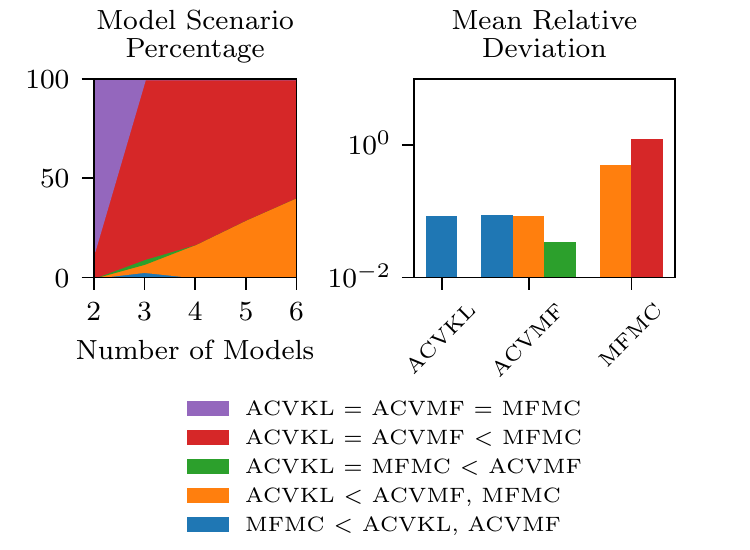}
\caption{Ordered Multifidelity}
\end{subfigure}
\begin{subfigure}{0.45\textwidth}
\vspace{10pt}
\includegraphics[scale=1]{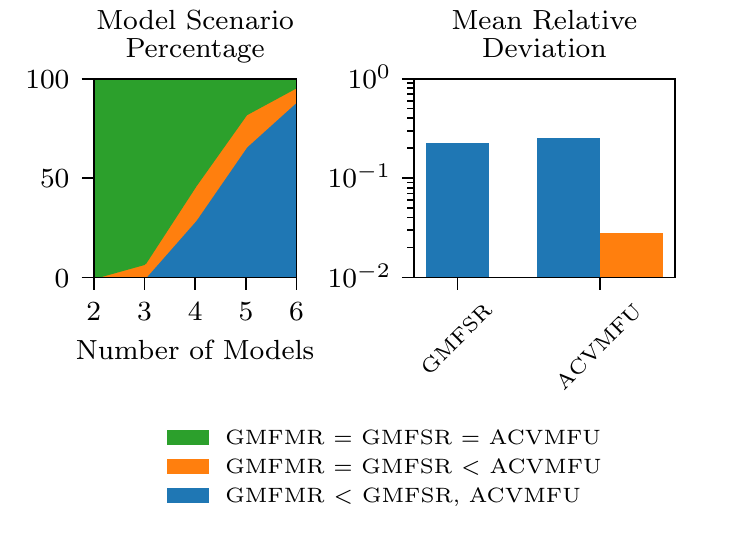}
\caption{Unordered Multifidelity}
\end{subfigure}
\caption{Relative performance of algorithms based on sampling strategy.  
}
\label{fig:other_sampling_strategies}
\end{figure}

The results for the independent samples strategy are shown in \fref{fig:independent_samples}.  In many model scenarios the algorithms perform equally.  For scenarios with 2 models, there is no difference between the three algorithms since only one recursion tree is possible.   For scenarios with 3 models, the GISSR and GISMR algorithms are also equivalent (because there are no possible recursion trees with depth greater than 3).    With increasing number of models, the more general search approaches start to outperform their more restrictive counterparts. This is due to the fact that the relative sizes of the optimization domains are diverging (see \fref{fig:num_recursions}).  Increased subdomain search results in better variance reduction.  On average, the variance reduction shows diminishing returns with size of subdomain.  This is illustrated by the mean relative deviation of GISMR over ACVIS compared to the improvement of GISMR over GISSR. Note that the relative deviation varies drastically between model scenarios.

The same basic trend is seen for the other three sampling strategies (see \fref{fig:other_sampling_strategies}): \ie with increasing number of models, the more general search approaches outperform their more restrictive counterparts.  Each of the other three sampling strategies have small notable differences as listed below.

\emph{Recursive difference strategy}:~ WRDIFF is not necessarily a subset of the GRDSR and so it slightly outperforms GRDSR on a small number of model scenarios.  Additionally, it should also be noted that MLMC outperforms the other recursive difference algorithms in 0.17\% of the model scenarios (not shown in the figure).  This is an artifact that occurs when three conditions are met: (1) the scenario has two models (2) the optimal $\alpha$ is very near 1, and (3) the non-optimal use of $\alpha=1$ by MLMC results in rounding of the number of samples in a favorable way.  In all other cases MLMC performs worse than the other $\alpha$-optimal algorithms.

\emph{Ordered multifidelity strategy}:~ The ACVKL includes the MFMC recursion tree but variance is calculated differently (numerical optimization versus analytically).  Thus MFMC can perform better than ACVKL in some model scenarios where sample rounding plays a part (similar to MLMC above).

\emph{Unordered multifidelity strategy}:~  GMFMR performs better than its more restrictive counterparts more frequently than any of the other generalized methods, perhaps indicating that the unordered multifidelity sampling strategy is particularly well suited for generalization.

\subsection{Comparison of Sampling Strategies}
\begin{figure}
\centering
\includegraphics[scale=1]{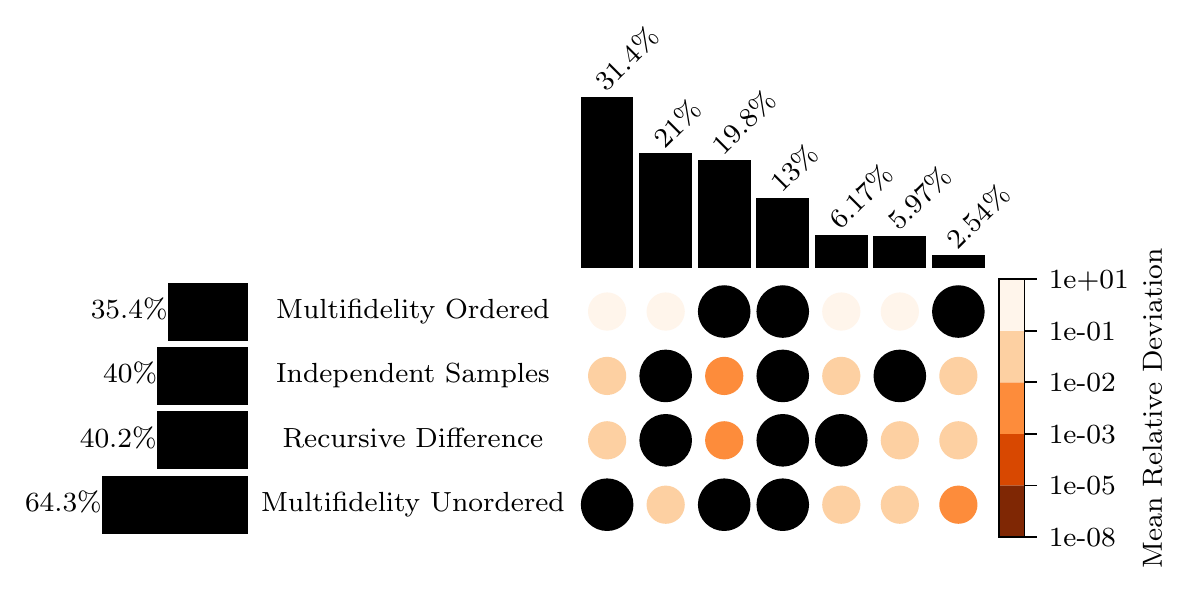}
\caption{Comparison of the performance of sampling strategies on the random model scenarios.  Columns of circles correspond to relative performance groups.  In each group the sampling strategy (or strategies) that performed best are indicated by a larger black circle.  Smaller colored circles represent the mean relative deviation of each strategy compared to the best strategy in that group.  The bars on the top indicate the percentage of model scenarios in each group.  The bars on the left indicate the percentage of model scenarios where each sampling strategy performed best. }
\label{fig:upset_no_ams}
\end{figure}

For each of the model scenarios, the best performing algorithm was found for each sampling strategy.  The sampling strategies are then compared to each other using those best algorithms. Similar to the previous subsection,  the model scenarios are grouped by the relative performance of the sampling strategies.  Specifically, they are grouped by which sampling strategy (or strategies) perform best on each model scenario.  The results of this analysis are illustrated in an upset plot (\fref{fig:upset_no_ams}).

The unordered multifidelity strategy performed best in most (64\%) model scenarios and outperformed all other strategies in around half of those scenarios (31.4\%).  The independent samples and recursive difference strategies often performed equally well, which makes sense considering how similar the strategies are.  When the ordered multifidelity method performed best, it was most frequently accompanied by unordered multifidelity, illustrating the cases where the unordered multifidelity model resulted in ordered models/recursion.  The exception is the small percentage of cases where MFMC performed best (see the previous section for a description of this artifact).  Based on the above, model scenarios tend to prefer either the multifidelty sampling strategies or independent samples/recursive difference strategies, without much overlap.  The overlap is defined by the model scenarios where all strategies perform equally.  This occurs when $M=1$ and the low-fidelity model is not well correlated with the high-fidelity model; all strategies produce a Monte Carlo result.

\begin{figure}
\centering
\includegraphics[scale=1]{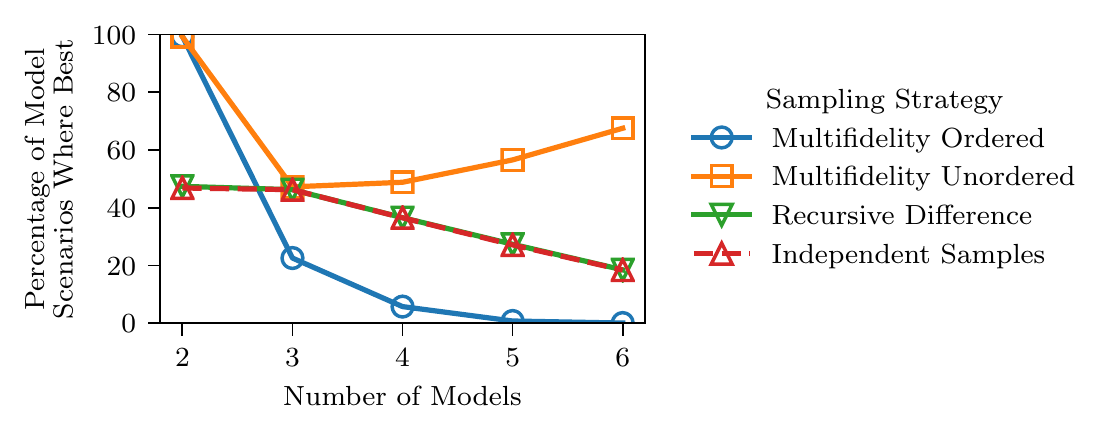}
\caption{The effect of model number on the relative performance of sampling strategies.}
\label{fig:category_v_m}
\end{figure}

The relative performance of the strategies are shown as a function of number of models in \fref{fig:category_v_m}.  The multifidelity strategies always perform best with 2 models.  As the number of models increases, the unordered multifidelity strategy performs increasingly well relative to the others (as might be expected considering \fref{fig:aggregate_performance_by_m}).  It is also worth noting that the ordered multifidelity strategy performs more poorly with increasing models because the chance of the models being \emph{correctly} ordered decreases.

\subsection{The Effect of Automatic Model Selection}
\begin{figure}
\centering
\includegraphics[scale=1.0]{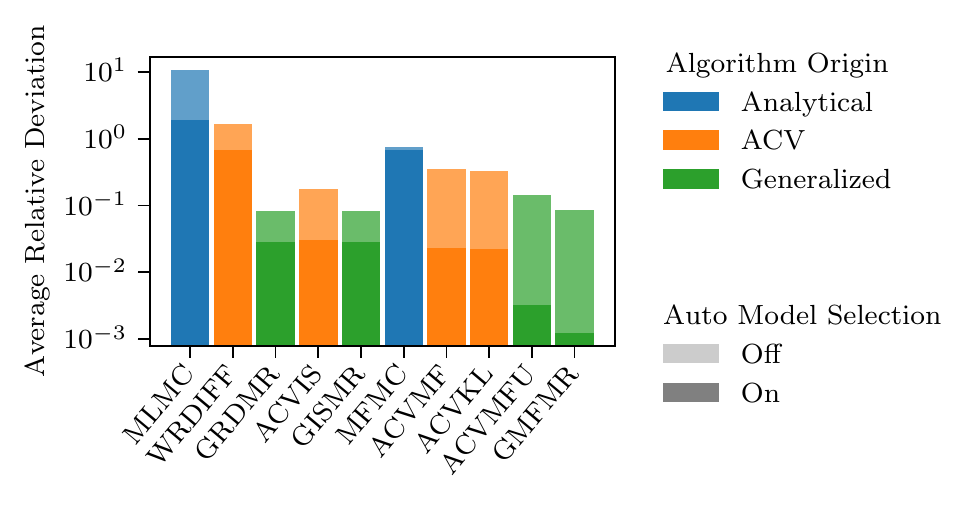}
\caption{Aggregate algorithm performance, including automatic model selection, on all random model scenarios.  The darker regions of each bar indicate automatic model section is on. The lighter region of each bar indicates automatic model selection is off.}
\label{fig:aggregate_ams}
\end{figure}

\fref{fig:aggregate_ams} illustrates the aggregate performance of several of the algorithms while not also considering automatic model selection:  $\mathcal{D}_{algs} = \{$MLMC, WRDIFF, GRDMR, ACVIS, GISMR, MFMC, ACVMF, ACVKL, ACVMFU, GMFMR, MLMC+AMS, WRDIFF+AMS, GRDMR+AMS, ACVIS+AMS, GISMR+AMS, MFMC+AMS, ACVMF+AMS, ACVKL+AMS, ACVMFU+AMS, GMFMR+AMS$\}$.  The trends are similar to those of the literature examples.  In contrast to \fref{fig:aggregate_performance_all}, GMFMR sees significantly better performance when automatic model selection is considered; it has a mean relative deviation of approximately 0.1\% which is 10 times smaller than any other algorithm (except for ACVMFU).

\begin{figure}
\centering
\includegraphics[scale=1]{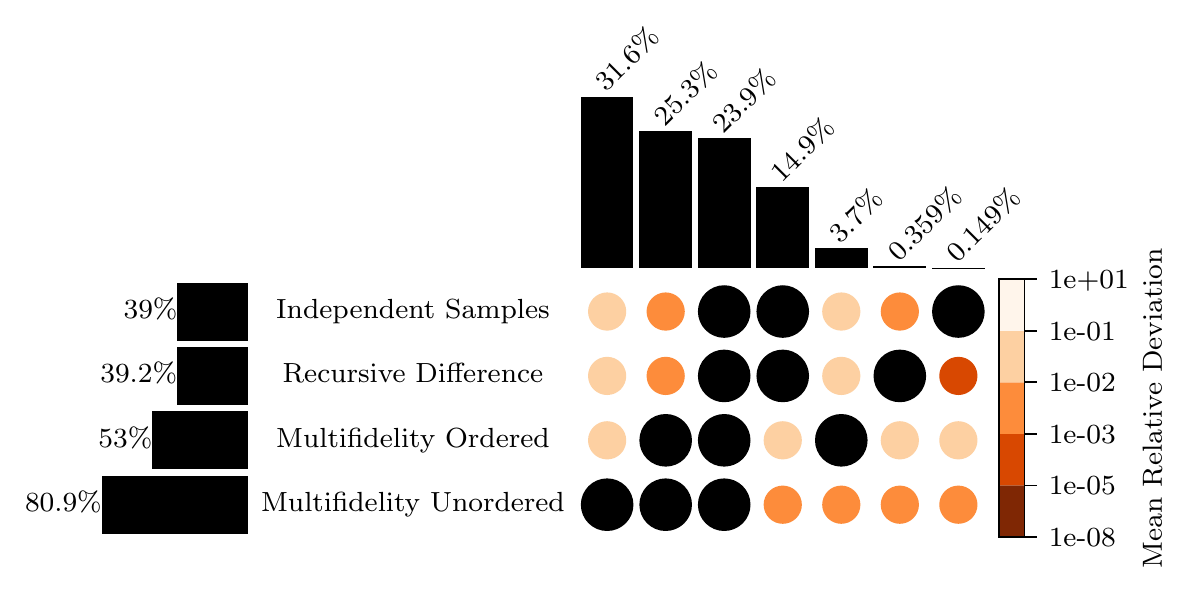}
\caption{Comparison of the performance of sampling strategies on the random model scenarios when automatic model selection is used.  Refer to \fref{fig:upset_no_ams} for full explanation of the formatting.}
\label{fig:upset_ams}
\end{figure}

The comparison of the sampling strategies is revisited using automatic model selection in \frefs{fig:upset_ams}{fig:category_v_m_ams}.  When including automatic model selection, multifidelity strategies more frequently result in the best variance reduction.  The unordered multifidelity strategy (GMFMR+AMS) performs best in more than 80\% of the random model scenarios when automatic model selection is used and averages less than 1\% deviation from the best algorithm in the scenarios where it did not perform best.  It also performs significantly better for all numbers of models.

\begin{figure}
\centering
\includegraphics[scale=1]{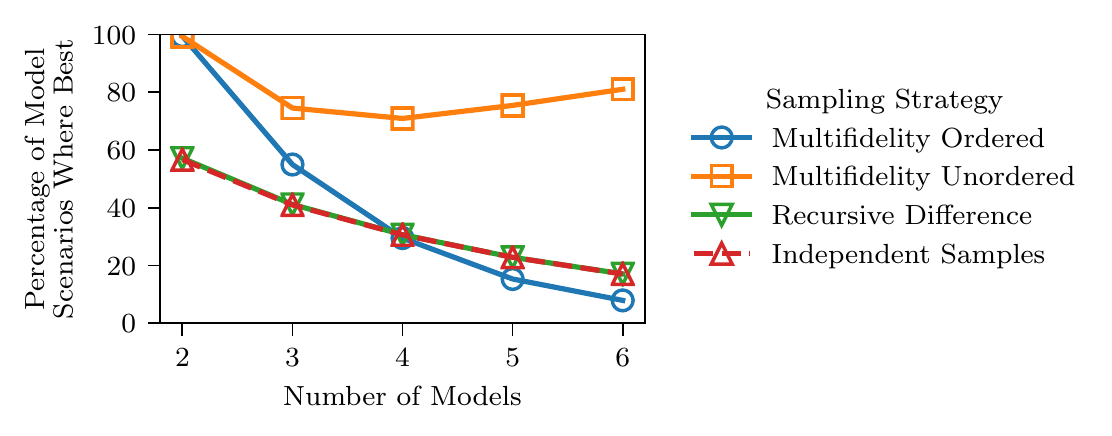}
\caption{The effect of model number on the relative performance of sampling strategies.}
\label{fig:category_v_m_ams}
\end{figure}

%% file: conclusion.tex
\section{Conclusion}\label{sec:conclusion}
In the current work, parametrically-defined estimators were explored as a means of generating better ACV estimators.  The formulation of a general form of the ACV estimator variance allowed for the creation of new parametrically-defined estimators based on arbitrary sample allocations.  Using this formulation, three sample allocation strategies (recursive difference, independent samples, and multifidelity) were generalized using the notion of tunable recursion.  The resulting parametrically-defined estimators were compared on a large set of model scenarios.  The primary findings are as follows:
\begin{itemize}
\item
The expansion of the effective optimization domain by parametrically-generated suboptimizations lead to ACVs with greater variance reduction.  

\item
Automatic model selection can produce sample allocations which are non-intuitive, which motivates its use in an automatic fashion rather than relying on human insight for construction of the estimators.

\item
The relaxation of ordering constraints frequently improved the ability of the numerical optimization to find improved sample allocations.  The effect of the ordering constraints diminishes when automatic model selection is used.

\item
In the majority of model scenarios studied in this work (including all literature scenarios) GMFMR with automatic model selection resulted in an estimator with lowest variance.  It becomes increasingly dominant as the number of models increases.
\end{itemize}

Because all of the investigated ACV estimators still rely on the restriction of the space of all possible sample allocations, there is no reason to believe that any of them contains the global optimum solution to the ACV optimization problem.  Thus, even more general solutions to the ACV optimization problem could be sought in future works.  In addition to providing explicit details on the estimators used in the current work, the Python package MXMCPy \cite{mxmcpy} has been released to facilitate the development and dissemination of new estimators throughout the community.

%% file: appendix.tex
\section*{Appendix}
\label{sec:appendix}

\subsection*{Dependence of $\Cov[\vec{\Delta}, \vec{\Delta}]$ and $\Cov[\vec{\Delta}, \hat Q_0]$ on $\mathcal{A}$}
The expressions for the dependence of $\Cov[\vec{\Delta}, \vec{\Delta}]$ and $\Cov[\vec{\Delta}, \hat Q_0]$ on sample allocation can be derived based on the covariance of MC estimators.  The covariance of two MC estimators is derived below based on the covariance of sums, sample independence, and the use of the model statistics.
\begin{equation}
\begin{split}
\Cov[\hat Q_i (z_j), \hat Q_k (z_\ell)] &= \Cov\left[ \frac{1}{N_j} \sum_m^{N_j} Q_i\left(z_j^{(m)} \right),  \frac{1}{N_\ell} \sum_n^{N_\ell} Q_k\left(z_\ell^{(n)} \right) \right]\\ 
&= \frac{1}{N_j N_\ell} \sum_m^{N_j} \sum_n^{N_\ell} \Cov\left[ Q_i\left(z_j^{(m)} \right), Q_k\left(z_\ell^{(n)} \right) \right]\\ 
&= \frac{1}{N_j N_\ell} \sum_m^{N_j} \sum_n^{N_\ell} \left\{\begin{array}{rl}
\Cov\left[ Q_i, Q_k \right] &: ~ z_j^{(m)} = z_\ell^{(n)}\\ 
0 &: ~ z_j^{(m)} \neq z_\ell^{(n)}\\ 
\end{array}\right\}\\
&= \frac{N_{j \cap \ell}}{N_jN_\ell}\Cov\left[Q_i , Q_k \right]
\label{eq:cov_mc_estimators}
\end{split}
\end{equation}

Focusing first on the elements of $\Cov[\vec{\Delta}, \hat Q_0]$, their general form can be derived from the definition of $\vec{\Delta}$, the covariance of sums, and \eref{eq:cov_mc_estimators}:
\begin{equation}
\begin{split}
\Cov[\vec{\Delta}_i, \hat Q_0] &= \Cov[\hat Q_i(z^*_i) - \hat Q_i(z_i), \hat Q_0(z_0)]\\
&= \Cov[\hat Q_i(z^*_i), \hat Q_0(z_0)] - \Cov[\hat Q_i(z_i), \hat Q_0(z_0)]\\
&= \frac{N_{i* \cap 0}}{N_{i*}N_0}\Cov[Q_i , Q_0 ] - \frac{N_{i \cap 0}}{N_{i}N_0}\Cov[Q_i , Q_0 ]\\
&= \left(\frac{N_{i* \cap 0}}{N_{i*}N_0} - \frac{N_{i \cap 0}}{N_{i}N_0} \right) \Cov[Q_i , Q_0 ]
\end{split}
\end{equation}
In the same fashion, the general form of the elements of $\Cov[\vec{\Delta}, \vec{\Delta}] $ is obtained:
\begin{equation}
\begin{split}
\Cov[\vec{\Delta}_i, \vec{\Delta}_j] =& \Cov[\hat Q_i(z^*_i) - \hat Q_i(z_i), \hat Q_j(z^*_j) - \hat Q_j(z_j)]\\
=& \Cov[\hat Q_i(z^*_i), \hat Q_j(z^*_j)] - \Cov[\hat Q_i(z^*_i), \hat Q_j(z_j)]\\
   & - \Cov[\hat Q_i(z_i), \hat Q_j(z^*_j)] + \Cov[\hat Q_i(z_i), \hat Q_j(z_j)]\\
=& \frac{N_{i* \cap j*}}{N_{i*}N_{j*}}\Cov[Q_i , Q_j ] - \frac{N_{i* \cap j}}{N_{i*}N_j}\Cov[Q_i , Q_j ]\\
   & - \frac{N_{i \cap j*}}{N_{i}N_{j*}}\Cov[Q_i , Q_j ] + \frac{N_{i \cap j}}{N_{i}N_j}\Cov[Q_i , Q_j ]\\
=& \left(\frac{N_{i* \cap j*}}{N_{i*}N_{j*}} - \frac{N_{i* \cap j}}{N_{i*}N_j}  - \frac{N_{i \cap j*}}{N_{i}N_{j*}}+ \frac{N_{i \cap j}}{N_{i}N_j} \right) \Cov[Q_i , Q_j ]
\end{split}
\end{equation}
Thus, the form of the relationship between $\Cov[\vec{\Delta}, \vec{\Delta}]$, $\Cov[\vec{\Delta}, \hat Q_0]$, and $\mathcal{A}$ is:
\begin{align}
\Cov[\vec{\Delta}, \vec{\Delta}] &= \mathbf{G}(\mathcal{A}) \circ \mathbf{C} \\
\Cov[\vec{\Delta}, \hat Q_0] &= \mathbf{g}(\mathcal{A}) \circ \mathbf{c}
\end{align}
with
\begin{align}
G_{ij} & = \frac{N_{i* \cap j*}}{N_{i*}N_{j*}} - \frac{N_{i* \cap j}}{N_{i*}N_j}  - \frac{N_{i \cap j*}}{N_{i}N_{j*}}+ \frac{N_{i \cap j}}{N_{i}N_j} \\
g_{i} & = \frac{N_{i* \cap 0}}{N_{i*}N_0} - \frac{N_{i \cap 0}}{N_{i}N_0}
\end{align}